\documentclass[fleqn,usenatbib]{mnras}

\usepackage{newtxtext,newtxmath}

\usepackage[T1]{fontenc}

\DeclareRobustCommand{\VAN}[3]{#2}
\let\VANthebibliography\thebibliography
\def\thebibliography{\DeclareRobustCommand{\VAN}[3]{##3}\VANthebibliography}

\usepackage{booktabs}   
\usepackage{graphicx}	
\usepackage{pgf}        
\usepackage{amsmath}	

\usepackage{siunitx}    
\usepackage{xspace}     %
\usepackage{ifthen}     %

\DeclareSIUnit\parsec{pc}
\DeclareSIUnit\lightyear{ly}
\DeclareSIUnit\year{yr}
\DeclareSIUnit\mag{mag}
\DeclareSIUnit\erg{erg}
\DeclareSIUnit\Msun{M_\odot}
\DeclareSIUnit\Lsun{L_\odot}
\DeclareSIUnit\dex{dex}
\DeclareSIUnit\percent{per cent}
\DeclareSIUnit\arcmin{arcmin}
\DeclareSIUnit\arcsec{arcsec}

\newcommand{\uncertainty}[3]{#1^{+#2}_{-#3}}
\newcommand\galaxyname[2]{#1\,#2}

\newcommand\ionized[2]{[\mathrm{#1}\,\textsc{#2}]}
\newcommand\OI[1][]{\ifthenelse{\equal{#1}{}}{\ionized{O}{i}}{\ionized{O}{i}\,\lambda#1}}
\newcommand\OIII[1][]{\ifthenelse{\equal{#1}{}}{\ionized{O}{iii}}{\ionized{O}{iii}\,\lambda#1}}
\newcommand\NII[1][]{\ifthenelse{\equal{#1}{}}{\ionized{N}{ii}}{\ionized{N}{ii}\,\lambda#1}}
\newcommand\SII[1][]{\ifthenelse{\equal{#1}{}}{\ionized{S}{ii}}{\ionized{S}{ii}\,\lambda#1}}
\newcommand\HA[1][]{\ifthenelse{\equal{#1}{}}{\mathrm{H}\,\alpha}{\mathrm{H}\,\alpha\,\lambda#1}}
\newcommand\HB[1][]{\ifthenelse{\equal{#1}{}}{\mathrm{H}\,\beta}{\mathrm{H}\,\beta\,\lambda#1}}

\newcommand\HII{H\,\textsc{ii}\xspace}
\newcommand\pn{\textsc{PN}\xspace}
\newcommand\pnlf{\textsc{PNLF}\xspace}

\newcommand\snr{\textsc{SNR}\xspace}
\newcommand\trgb{\textsc{TRGB}\xspace}
\newcommand\sbf{\textsc{SBF}\xspace}
\newcommand\ifu{\textsc{IFU}\xspace}

\usepackage{pgfkeys}

\newcommand{\setvalue}[1]{\pgfkeys{/variables/#1}}
\newcommand{\getvalue}[1]{\pgfkeysvalueof{/variables/#1}}
\newcommand{\declare}[1]{%
 \pgfkeys{
  /variables/#1.is family,
  /variables/#1.unknown/.style = {\pgfkeyscurrentpath/\pgfkeyscurrentname/.initial = ##1}
 }%
}

\declare{Mpc/}
\declare{mag/}

\setvalue{mag/IC5332 = \uncertainty{29.73}{0.10}{0.20}}
\setvalue{mag/NGC0628 = \uncertainty{29.89}{0.06}{0.09}}
\setvalue{mag/NGC1087 = \uncertainty{31.05}{0.10}{0.24}}
\setvalue{mag/NGC1300 = \uncertainty{32.06}{0.08}{0.12}}
\setvalue{mag/NGC1365 = \uncertainty{31.22}{0.08}{0.14}}
\setvalue{mag/NGC1385 = \uncertainty{29.96}{0.14}{0.32}}
\setvalue{mag/NGC1433 = \uncertainty{31.39}{0.04}{0.07}}
\setvalue{mag/NGC1512 = \uncertainty{31.27}{0.07}{0.11}}
\setvalue{mag/NGC1566 = \uncertainty{31.13}{0.08}{0.17}}
\setvalue{mag/NGC1672 = \uncertainty{30.99}{0.09}{0.23}}
\setvalue{mag/NGC2835 = \uncertainty{30.57}{0.08}{0.17}}
\setvalue{mag/NGC3351 = \uncertainty{30.36}{0.06}{0.08}}
\setvalue{mag/NGC3627 = \uncertainty{30.18}{0.08}{0.15}}
\setvalue{mag/NGC4254 = \uncertainty{29.97}{0.09}{0.20}}
\setvalue{mag/NGC4303 = \uncertainty{30.65}{0.10}{0.26}}
\setvalue{mag/NGC4321 = \uncertainty{31.10}{0.06}{0.10}}
\setvalue{mag/NGC4535 = \uncertainty{31.43}{0.06}{0.09}}
\setvalue{mag/NGC5068 = \uncertainty{28.46}{0.11}{0.26}}
\setvalue{mag/NGC7496 = \uncertainty{31.64}{0.09}{0.19}}
 
\setvalue{Mpc/IC5332 = \uncertainty{8.84}{0.39}{0.82}}
\setvalue{Mpc/NGC0628 = \uncertainty{9.52}{0.26}{0.41}}
\setvalue{Mpc/NGC1087 = \uncertainty{16.25}{0.74}{1.79}}
\setvalue{Mpc/NGC1300 = \uncertainty{25.77}{0.90}{1.42}}
\setvalue{Mpc/NGC1365 = \uncertainty{17.53}{0.66}{1.16}}
\setvalue{Mpc/NGC1385 = \uncertainty{9.81}{0.63}{1.46}}
\setvalue{Mpc/NGC1433 = \uncertainty{18.94}{0.39}{0.56}}
\setvalue{Mpc/NGC1512 = \uncertainty{17.93}{0.53}{0.88}}
\setvalue{Mpc/NGC1566 = \uncertainty{16.84}{0.60}{1.29}}
\setvalue{Mpc/NGC1672 = \uncertainty{15.80}{0.68}{1.68}}
\setvalue{Mpc/NGC2835 = \uncertainty{13.03}{0.46}{1.04}}
\setvalue{Mpc/NGC3351 = \uncertainty{11.80}{0.31}{0.43}}
\setvalue{Mpc/NGC3627 = \uncertainty{10.88}{0.39}{0.77}}
\setvalue{Mpc/NGC4254 = \uncertainty{9.86}{0.42}{0.91}}
\setvalue{Mpc/NGC4303 = \uncertainty{13.49}{0.64}{1.60}}
\setvalue{Mpc/NGC4321 = \uncertainty{16.62}{0.46}{0.74}}
\setvalue{Mpc/NGC4535 = \uncertainty{19.29}{0.56}{0.82}}
\setvalue{Mpc/NGC5068 = \uncertainty{4.93}{0.24}{0.59}}
\setvalue{Mpc/NGC7496 = \uncertainty{21.31}{0.89}{1.89}}

\title[PNLF distances for 19 nearby galaxies observed by PHANGS]{Planetary Nebula Luminosity Function distances for 19 galaxies observed by PHANGS--MUSE}

\author[F. Scheuermann et al.]{Fabian Scheuermann$^{1}$\thanks{E-mail: f.scheuermann@uni-heidelberg.de},
Kathryn Kreckel$^{1}$,
Gagandeep S. Anand$^{2,3}$,
Guillermo A. Blanc$^{4,5}$,
\newauthor
Enrico Congiu$^{5}$,
Francesco Santoro$^{6}$,
Schuyler D. Van Dyk$^{7}$, 
Ashley~T.~Barnes$^{8}$,
\newauthor
Frank Bigiel$^{8}$,
Simon C. O. Glover$^{9}$,
Brent Groves$^{10}$,
Ralf S. Klessen$^{9,11}$,
\newauthor
J. M. Diederik Kruijssen$^{1}$,
Erik Rosolowsky$^{12}$,
Eva Schinnerer$^{6}$,
Andreas Schruba$^{13}$,
\newauthor
Elizabeth J. Watkins$^{1}$,
Thomas G. Williams$^{6}$
\\
Affiliations are listed at the end of the paper.
}

\date{Accepted XXX. Received YYY; in original form ZZZ}

\pubyear{2022}

\begin{document}
\label{firstpage}
\pagerange{\pageref{firstpage}--\pageref{lastpage}}
\maketitle

\begin{abstract}
We provide new planetary nebula luminosity function (\pnlf) distances to 19 nearby spiral galaxies that were observed with VLT/MUSE by the PHANGS collaboration. Emission line ratios are used to separate planetary nebulae (\pn{}e) from other bright $\OIII$ emitting sources like compact supernovae remnants (\snr{}s) or \HII regions. While many studies have used narrowband imaging for this purpose, the detailed spectral line information provided by integral field unit (\ifu) spectroscopy grants a more robust way of categorising different $\OIII$ emitters. We investigate the effects of \snr contamination on the \pnlf and find that we would fail to classify all objects correctly, when limited to the same data narrowband imaging provides. However, the few misclassified objects usually do not fall on the bright end of the luminosity function, and only in three cases does the distance change by more than $1\sigma$. We find generally good agreement with literature values from other methods. Using metallicity constraints that have also been derived from the same \ifu data, we revisit the \pnlf zero point calibration. Over a range of $8.34 < 12+\log(\mathrm{O}/\mathrm{H})<8.59$, our sample is consistent with a constant zero point and yields a value of $M^*=\uncertainty{-4.542}{0.103}{0.059}\, \si{\mag}$, within $1\sigma$ of other literature values. MUSE pushes the limits of \pnlf studies and makes galaxies beyond $\SI{20}{\mega\parsec}$ accessible for this kind of analysis. This approach to the \pnlf shows great promise for leveraging existing archival \ifu data on nearby galaxies.  
\end{abstract}





\defcitealias{1981ApJ...244..780B}{Branch+81}
\defcitealias{1981ApJ...248..408D}{de Vaucouleurs+81}
\defcitealias{1981ApJS...45..541P}{Pedreros+81}
\defcitealias{1981PASP...93...36D}{de Vaucouleurs+81}
\defcitealias{1984A&AS...56..381B}{Bottinelli+84}
\defcitealias{1985A&A...153..125G}{Giraud+85}
\defcitealias{1985A&AS...59...43B}{Bottinelli+85}
\defcitealias{1985Natur.318...25B}{Bartel+85}
\defcitealias{1986A&A...156..157B}{Bottinelli+86}
\defcitealias{1988NBGC.C....0000T}{Tully+88}
\defcitealias{1992ApJ...395..366S}{Schmidt+92}
\defcitealias{1992ApJS...81..413M}{Persic+92}
\defcitealias{1994ApJ...432...42S}{Schmidt+94}
\defcitealias{1994ApJ...433...19S}{Boffi+94}
\defcitealias{1996A&AS..119..499S}{Sharina+96}
\defcitealias{1996AJ....111.2280S}{Sohn+96}
\defcitealias{1996ApJ...457..500H}{Hoeflich+96}
\defcitealias{1996ApJ...465L..83R}{Ruiz-Lapuente+96}
\defcitealias{1996AstL...22...71Z}{Zasov+96}
\defcitealias{1997A&A...323...14S}{Schoeniger+97}
\defcitealias{1997ApJS..108..417Y}{Yasuda+97}
\defcitealias{1997ApJS..109..333W}{Willick+97}
\defcitealias{1997AstL...23..644G}{Georgiev+97}
\defcitealias{2000A&A...355..835E}{Ekholm+00}
\defcitealias{2002A&A...389...19P}{Paturel+02}
\defcitealias{2002A&A...393...57T}{Terry+02}
\defcitealias{2002AJ....123..207D}{Dolphin+02}
\defcitealias{2002ApJ...565..681R}{Russell+02}
\defcitealias{2002ApJ...577...31C}{Ciardullo+02}
\defcitealias{2003A&A...411..361K}{Kanbur+03}
\defcitealias{2003ApJ...591..301B}{Bartel+03}
\defcitealias{2004ApJ...608...42S}{Sakai+04}
\defcitealias{2005ApJ...624..532R}{Reindl+05}
\defcitealias{2005MNRAS.359..906H}{Hendry+05}
\defcitealias{2006A&A...452..423P}{Paturel+06}
\defcitealias{2006ApJ...645..841N}{Nugent+06}
\defcitealias{2006ApJS..165..108S}{Saha+06}
\defcitealias{2006PASP..118..351V}{Van Dyk+06}
\defcitealias{2007A&A...465...71T}{Theureau+07}
\defcitealias{2007ApJ...659..122J}{Jha+07}
\defcitealias{2007ApJ...661..815R}{Rizzi+07}
\defcitealias{2008A&ARv..15..289T}{Tammann+08}
\defcitealias{2008ApJ...683..630H}{Herrmann+08}
\defcitealias{2008MNRAS.389.1577T}{Takanashi+08}
\defcitealias{2009AJ....138..323T}{Tully+09}
\defcitealias{2009AJ....138..332J}{Jacobs+09}
\defcitealias{2009ApJ...694.1067P}{Poznanski+09}
\defcitealias{2009ApJ...697..996M}{Mould+09}
\defcitealias{2009ApJS..182..474S}{Springob+09}
\defcitealias{2010ApJ...715..833O}{Olivares+10}
\defcitealias{2010ApJ...716..712A}{Amanullah+10}
\defcitealias{2011A&A...532A.104N}{Nasonova+11}
\defcitealias{2011AJ....141...19B}{Dhawan+11}
\defcitealias{2011ApJ...731..120M}{Mandel+11}
\defcitealias{2011ApJ...736...76R}{Roy+11}
\defcitealias{2012ApJ...749..174C}{Courtois+12}
\defcitealias{2012ApJ...758L..12S}{Sorce+12}
\defcitealias{2013AJ....145..101K}{Karachentsev+13}
\defcitealias{2013AJ....146...86T}{Tully+13}
\defcitealias{2013ApJ...771...88L}{Lagattuta+13}
\defcitealias{2013ApJ...773...13L}{Lee+13}
\defcitealias{2013ApJ...773...53F}{Folatelli+13}
\defcitealias{2014AJ....148....1Z}{Zhang+14}
\defcitealias{2014AJ....148..107R}{Rodríguez+14}
\defcitealias{2014ApJ...782...98B}{Bose+14}
\defcitealias{2014ApJ...792...52J}{Jang+14}
\defcitealias{2014MNRAS.444..527S}{Sorce+14}
\defcitealias{2015A&A...580L..15P}{Polshaw+15}
\defcitealias{2015ApJ...799..215P}{Pejcha+15}
\defcitealias{2015ApJ...806..195V}{Van Dyk+15}
\defcitealias{2015MNRAS.448.2312B}{Barbarino+15}
\defcitealias{2016AJ....152...50T}{Tully+16}
\defcitealias{2016ApJ...822....6D}{Dhungana+16}
\defcitealias{2016ApJ...826...56R}{Riess+16}
\defcitealias{2017AJ....154...51M}{McQuinn+17}
\defcitealias{2017ApJ...834..174K}{Kreckel+17}
\defcitealias{2017ApJ...836...74J}{Jang+17}
\defcitealias{2017ApJ...841..127M}{Müller+17}
\defcitealias{2017ApJ...843...16K}{Kourkchi+17}
\defcitealias{2017ApJ...846...58H}{Hoeflich+17}
\defcitealias{2017ApJ...850..207S}{Shaya+17}
\defcitealias{2017MNRAS.469L.113K}{Karachentsev+17}
\defcitealias{2018ApJ...852...60J}{Jang+18}
\defcitealias{2018ApJS..235...23S}{Sabbi+18}
\defcitealias{2020AJ....159...67K}{Kourkchi+20}
\defcitealias{Freedman+2001}{Freedman+01}
\defcitealias{Ciardullo+2002}{Ciardullo+02}
\defcitealias{Herrmann+2008}{Herrmann+08}
\defcitealias{Kreckel+2017}{Kreckel+17}
\defcitealias{Anand+2021}{Anand+21}
\defcitealias{Roth+2021}{Roth+21}

\section{Introduction}

Observations are the primary way to gather data in astronomy, but not all information is directly accessible. The measurement of some properties, like the intrinsic luminosity or the physical size of galaxies or stars depend on their distances. It is therefore fundamental for our understanding of the Universe to know those distances, but measuring them is a delicate task. Cosmic distances span many orders of magnitude, which prohibits the use of a single distance measure and instead necessitates a combination of techniques -- the so-called \emph{cosmic distance ladder}. Unfortunately, flaws in the different rungs can propagate into large uncertainties for the most distant objects \citep{Bernal+2016,Freedman+2021}. It is therefore desirable to find techniques that are accurate and easily applicable to a large number of objects. 

For nearby galaxies, the \emph{tip of the red giant branch} \citep[\trgb, e.g.~][]{Lee+1993}, \emph{Cepheids} \citep[e.g.~][]{Freedman+2001}, \emph{Type Ia supernovae} \citep[SNe Ia, e.g.~][]{Riess+1996} and \emph{Surface Brightness Fluctuations} \citep[SBF, e.g.~][]{Tonry+2001} are among the premier techniques used to obtain precise distances, but redundant methods are essential to check for systematic differences. Another method that can achieve similar precision is the \emph{planetary nebulae luminosity function} (\pnlf). \citet{Ford+1978} were among the first to realise that planetary nebulae  (\pn{}e) can be utilised to determine the distance to their host galaxy. A \pn is the last breath of a dying intermediate mass star \citep[\SIrange{2}{8}{\Msun};][]{Kwok+2005}. During the final stages of its life, the star will expels its outer layers, which will then be ionised by the UV radiation of the exposed hot core. The low densities of the rarefied gas, enables a wealth of collisionally excited emission lines \citep{Osterbrock+2006}. Even though many stars will go through this phase at the end of their lives, the actual number of \pn{}e that we observe is rather small, because they only last for a few tens of thousands of years \citep{Buzzoni+2006}.

Since the maximum absolute magnitude of all \pn{}e is roughly constant across all galaxies, they can be used as standard candles. \citet{Jacoby+1989a} and \citet{Ciardullo+1989a} formulated an empirical luminosity function that made use of this luminosity cut-off and in doing so, laid the groundwork for how we measure distances with \pn{}e to the present day. Since then, the \pnlf has been established as a reliable method for obtaining distances to nearby galaxies up to ${\sim}\SI{20}{\mega\parsec}$, with a precision better than $\SI{10}{\percent}$ \citep[e.g.,][]{Ciardullo+1993,Ferrarese+2000a,Ciardullo+2002,Feldmeier+2007}. 

The \pnlf also plays a special role in the cosmic distance ladder. To the best of our knowledge \citep[though still lacking a good theoretical foundation,][]{Ciardullo+2013}, the bright end cutoff of the \pnlf is invariant to changes in the local environment or the metallicity and is applicable to both spiral \citep{Feldmeier+1997} and elliptical galaxies \citep{Ciardullo+1989b}. This makes this method useful for testing and comparing other methods. Beyond their use as a distance indicator, \pn{}e are also used to measure kinematics in the outer halo where measuring stellar kinematics directly is difficult \citep{Hartke+2017}.

The greatest challenge in measuring the \pnlf distance is to compile a clean catalogue of \pn{}e. Due to their strong $\OIII[5007]$  emission \citep{Schoenberner+2010}, they are easily detectable, but it can be difficult to discriminate them from other nebular objects like \HII regions or supernova remnants (\snr). \citet{Baldwin+1981} showed that emission line ratios can be used to make this discrimination. To this end, narrowband imaging with a combination of an $\OIII$ on-band and a wider off-band filter together with a filter centred around $\HA$, are used to identify \pn{}e.

In recent years, the emergence of powerful \emph{integral field unit} (\ifu) spectroscopy instruments has revolutionised the field  \citep{Kreckel+2017,Spriggs+2020, Roth+2021}. Their fields of view are large enough to detect a sufficient number of \pn{}e in order to measure the \pnlf. \pn{}e trace the underlying stellar population and hence we expect many more \pn{}e in the denser central regions. This means that we can potentially find more objects in a smaller field of view. However, the strong stellar continuum in this part of the galaxy also makes it difficult to detect them with narrowband imaging. With the full spectral information from the \ifu, we can remove the continuum and search for \pn{}e in regions that were previously inaccessible.

We use data that were observed for \textbf{P}hysics at \textbf{H}igh \textbf{A}ngular resolution in \textbf{N}earby \textbf{G}alaxie\textbf{S} (PHANGS\footnote{\url{http://www.phangs.org}}). This is a collaboration aimed at studying the baryon cycle within galaxies at high spatial resolution, sufficient to isolate and characterise individual molecular clouds and \HII regions. One of the pillars of this project are optical \ifu spectroscopy observations with the Multi Unit Spectroscopic Explorer \citep[MUSE;][]{Bacon+2010} at the Very Large Telescope (VLT). The PHANGS--MUSE sample consists of 19 nearby spiral galaxies that are all roughly face-on, ranging in mass from $\log_{10} (M/\si{\Msun}) = 9.4$ to $10.99$ and star formation rate from  $\log_{10} (\mathrm{SFR}/\si{\Msun \per \year}) = -0.56$ to $0.88$.

This paper has three objectives. The first aim is to test how susceptible narrowband imaging is to misclassifying \snr{}s as \pn{}e. \citet{Kreckel+2017} found that, in the case of the nearby spiral galaxy \galaxyname{NGC}{0628}, contamination with \snr{}s can bias the measured distance. Meanwhile, \citet{Davis+2018} could not find such problems for \galaxyname{M}{31} and \galaxyname{M}{33}. The lingering question is whether \galaxyname{NGC}{0628} was just an anomaly or if this issue compromises other galaxies as well. 

The second goal is to quantify the applicability of \ifu surveys for \pn studies. Past \pn studies required special observations that were taken for the sole purpose of finding \pn{}e. Because \ifu studies cover the relevant emission lines anyway, this opens up the possibility of measuring the \pnlf as a `bonus'. The gain in versatility comes at a price: the fields of view are usually narrower. Here we try to quantify whether \ifu surveys can compensate the smaller fields of view with a gain in spectral information and explorer the limitations to derive \pnlf-based distances from \ifu observations.

The last objective is to provide new and precise distance measurements for the 19 nearby galaxies observed by the PHANGS--MUSE survey, some of which did not have good distance estimates before. In Section~\ref{sec:findingPN}, we present the data that is used for this study and describe our process for identifying \pn{}e. In Section~\ref{sec:pnlf}, we fit the observed data to the \pnlf and in Section~\ref{sec:results} and~\ref{sec:discussion} we present and discuss the results. We conclude in section~\ref{sec:conclusion}.

\section{Finding planetary nebulae}\label{sec:findingPN}

\begin{table*}
\centering
\caption{Properties of the galaxies in the PHANGS--MUSE sample}
\begin{tabular}{
    l
    l
    r
    r
    S[table-format=2.2]@{\,\( \pm \)\,}
    S[table-format=1.2]
    S[table-format=2.2]
    S[table-format=3.2]
    S[table-format=1.2]
    S[table-format=2.2]
    S[table-format=1.3]
    l
    }
\toprule\toprule
{Name} & {Type} & \multicolumn{1}{c}{R.A.} & \multicolumn{1}{c}{Dec.} & \multicolumn{2}{c}{$(m-M)^\mathrm{a}$} & {$i{}^\mathrm{b}$} & {PA${}^\mathrm{b}$} & {$r_{25}$} & {$12+\log(\mathrm{O}/\mathrm{H})^\mathrm{c}$} & {$E(B-V)^\mathrm{d}$} & {FWHM${}^\mathrm{e}$} \\
 &  & \multicolumn{1}{c}{(J2000)} & \multicolumn{1}{c}{(J2000)} & \multicolumn{2}{c}{$\si{\mag}$} & {deg} & {deg} & {arcmin} &  &  & {arcsec}  \\
\midrule
\galaxyname{IC}{5332} & SABc & 23h34m27.49s & $-$36d06m03.89s & 29.77 & 0.10 & 26.9 & 74.4 & 3.03 & 8.38 & 0.015 & $0.72^\mathrm{AO}$ \\
\galaxyname{NGC}{0628} & Sc & 01h36m41.73s & $+$15d47m01.11s & 29.96 & 0.14 & 8.9 & 20.7 & 4.94 & 8.50 & 0.062 & $0.73^\mathrm{AO}$ \\
\galaxyname{NGC}{1087} & Sc & 02h46m25.18s & $-$00d29m55.38s & 31.00 & 0.30 & 42.9 & 359.1 & 1.49 & 8.37 & 0.030 & $0.74^\mathrm{AO}$ \\
\galaxyname{NGC}{1300} & Sbc & 03h19m41.00s & $-$19d24m40.01s & 31.39 & 0.33 & 31.8 & 278.0 & 2.97 & 8.52 & 0.026 & $0.63$ \\
\galaxyname{NGC}{1365} & Sb & 03h33m36.36s & $-$36d08m25.45s & 31.46 & 0.09 & 55.4 & 201.1 & 6.01 & 8.54 & 0.018 & $0.82^\mathrm{AO}$ \\
\galaxyname{NGC}{1385} & Sc & 03h37m28.56s & $-$24d30m04.18s & 31.18 & 0.33 & 44.0 & 181.3 & 1.70 & 8.40 & 0.018 & $0.49$ \\
\galaxyname{NGC}{1433} & SBa & 03h42m01.49s & $-$47d13m18.99s & 29.78 & 0.49 & 28.6 & 199.7 & 3.10 & 8.55 & 0.008 & $0.65$ \\
\galaxyname{NGC}{1512} & Sa & 04h03m54.14s & $-$43d20m55.41s & 30.28 & 0.33 & 42.5 & 261.9 & 4.22 & 8.56 & 0.009 & $0.8^\mathrm{AO}$ \\
\galaxyname{NGC}{1566} & SABb & 04h20m00.38s & $-$54d56m16.84s & 31.24 & 0.25 & 29.5 & 214.7 & 3.61 & 8.55 & 0.008 & $0.64$ \\
\galaxyname{NGC}{1672} & Sb & 04h45m42.49s & $-$59d14m50.13s & 31.44 & 0.33 & 42.6 & 134.3 & 3.08 & 8.54 & 0.021 & $0.72^\mathrm{AO}$ \\
\galaxyname{NGC}{2835} & Sc & 09h17m52.91s & $-$22d21m16.84s & 30.44 & 0.17 & 41.3 & 1.0 & 3.21 & 8.38 & 0.089 & $0.85^\mathrm{AO}$ \\
\galaxyname{NGC}{3351} & Sb & 10h43m57.76s & $+$11d42m13.21s & 29.99 & 0.07 & 45.1 & 193.2 & 3.61 & 8.59 & 0.024 & $0.74^\mathrm{AO}$ \\
\galaxyname{NGC}{3627} & Sb & 11h20m15.00s & $+$12d59m29.40s & 30.27 & 0.09 & 57.3 & 173.1 & 5.14 & 8.55 & 0.037 & $0.77^\mathrm{AO}$ \\
\galaxyname{NGC}{4254} & Sc & 12h18m49.63s & $+$14d24m59.08s & 30.59 & 0.33 & 34.4 & 68.1 & 2.52 & 8.53 & 0.035 & $0.58$ \\
\galaxyname{NGC}{4303} & Sbc & 12h21m54.93s & $+$04d28m25.48s & 31.15 & 0.39 & 23.5 & 312.4 & 3.44 & 8.56 & 0.020 & $0.58$ \\
\galaxyname{NGC}{4321} & SABb & 12h22m54.93s & $+$15d49m20.29s & 30.91 & 0.07 & 38.5 & 156.2 & 3.05 & 8.56 & 0.023 & $0.64$ \\
\galaxyname{NGC}{4535} & Sc & 12h34m20.30s & $+$08d11m52.70s & 30.99 & 0.05 & 44.7 & 179.7 & 4.07 & 8.55 & 0.017 & $0.44$ \\
\galaxyname{NGC}{5068} & Sc & 13h18m54.74s & $-$21d02m19.48s & 28.58 & 0.09 & 35.7 & 342.4 & 3.74 & 8.34 & 0.091 & $0.73^\mathrm{AO}$ \\
\galaxyname{NGC}{7496} & Sb & 23h09m47.29s & $-$43d25m40.26s & 31.36 & 0.33 & 35.9 & 193.7 & 1.67 & 8.51 & 0.008 & $0.79$ \\
\bottomrule
\multicolumn{11}{l}{Adopted from the PHANGS sample table \citep[v1p6, see][]{Leroy+2021b}.} \\
\multicolumn{11}{l}{$^\mathrm{a}$ Curated by \citet{Anand+2021}. For \galaxyname{NGC}{1433} and \galaxyname{NGC}{1512} the distances are from \citet{2018ApJS..235...23S}.} \\
\multicolumn{11}{l}{$^\mathrm{b}$ Inclination and position angle from \citet{Lang+2020}.} \\
\multicolumn{11}{l}{$^\mathrm{c}$ At the mean positions of the \pn{}e with gradients from  \citet{Kreckel+2020} and \citet{Santoro+2021}.} \\
\multicolumn{11}{l}{$^\mathrm{d}$ Galactic foreground extinction from \citet{Schlafly+2011}.}\\
\multicolumn{11}{l}{$^\mathrm{e}$ Average FWHM of the PSF across all pointings. Galaxies that were observed with adaptive optics are marked with AO.}
\end{tabular}
\label{tbl:sample}
\end{table*}

We begin by compiling a clean catalogue of \pn{} candidates followed by a second catalogue with potential \snr{} contaminants that might be misclassified in narrowband studies.

\subsection{Data}

We use the full sample of 19 nearby spiral galaxies (see Table~\ref{tbl:sample}) that were observed with VLT/MUSE for the PHANGS--MUSE large observing program \citep[PI: Schinnerer;][]{Emsellem+2021}. The wide-field mode of MUSE has a field of view of $1\times\SI{1}{\arcmin\squared}$ and only covers part of the targeted galaxies. Between 3 and 15 pointings were taken to sample the bulk of the star-forming disk of each target. The observations cover a wavelength range from \SIrange{4800}{9300}{\angstrom}, with a typical spectral resolution of $\SI{2.75}{\angstrom}$. Each pixel covers $0.2\times\SI{0.2}{\arcsec\squared}$ with an average spatial resolution of $\SI{0.72}{\arcsec}$ ($\SI{67}{\parsec}$). 

The data reduction was performed by the PHANGS--MUSE team, using the MUSE data processing pipeline by \citet{Weilbacher+2020}, and is described in \citet{Emsellem+2021}. They produced reduced and mosaicked spectral cubes that form the base for further data analysis pipeline (DAP) products which we use for our analysis. This includes emission line maps that are extracted by fitting Gaussian profiles along with the stellar continuum. We use those maps for the $\HB[4861]$, $\HA[6562]$, $\NII[6583]$, $\SII[6716]$ and $\SII[6731]$ emission lines. We also use the $\OIII[5007]$ line maps, which are not tied to any of the other lines during the fit, for PN source detection. For our analysis we use the native resolution DAP maps from the public data release 1.0.

However, this procedure is not suitable to obtain accurate $\OIII[5007]$ fluxes. $\OIII$ emission is mostly associated with certain objects like \pn{}e, \snr{}s and \HII regions, and most parts of a galaxy will have little to no emission. By fitting a Gaussian profile in a region without emission, we create a positive bias. To circumvent this issue, we calculate the non-parametric emission-line moment for $\OIII[5007]$ following the procedure employed in the MaNGA data analysis pipeline \citep{Belfiore+2019,Westfall+2019}. We use the velocity from the $\HA$ line to calculate the expected rest-frame wavelength of the $\OIII$ line and sum the flux over 17 spectral bins, corresponding to a wavelength range of \SIrange[]{4998}{5018}{\angstrom}. This range is wide enough to ensure that we always capture the $\OIII$ line, even in the case that the velocity from the $\HA$ is off. In a similar way, the continuum is measured in the range from \SIrange{4978}{4998}{\angstrom} and \SIrange{5018}{5038}{\angstrom} with three sigma clipping and then subtracted from the line flux.

In order to measure the flux of a point source like a \pn, one needs to know the shape of the point spread function (PSF). For our MUSE data, it is best parameterised by a Moffat profile \citep{Bacon+2017,Fusco+2020}
\begin{equation}\label{eq:Moffat}
    p(r) = A \left[ 1+\left(\frac{r^2}{\gamma^2}\right) \right]^{-\alpha}.
\end{equation}
A fixed power index of $\alpha = 2.3$ is used for galaxies with adaptive optics (AO) observations and $\alpha=2.8$ for the remaining galaxies (both values were provided by ESO). The core width $\gamma$ varies from pointing to pointing and is related to the full width half maximum via $\mathrm{FWHM} = 2 \gamma \sqrt{2^{1/\alpha}-1}$. The latter is usually measured from bright point-like objects. However since many pointings in our data do not have such objects, we use a different approach where we measure the FWHM with the help of a reference image with a known PSF \citep{Emsellem+2021}. Unfortunately the derived FWHM can have a rather large uncertainties of up to $\SI{0.1}{\arcsec}$. The FWHM is measured at a reference wavelength of $\SI{6483.58}{\angstrom}$ and it decreases linearly with increasing wavelength. This can be approximated with  
\begin{equation}
\mathrm{FWHM} (\lambda) = \mathrm{FWHM}_\mathrm{ref} - \SI{3e-5}{\arcsec\per\angstrom}(\lambda-\SI{6483.58}{\angstrom}),
\end{equation}
where $\mathrm{FWHM}_\text{ref}$ is measured at the reference wavelength.

\subsection{Source detection and photometry}
\label{sec:detection}

\begin{figure*}
\centering
\includegraphics[width=\textwidth]{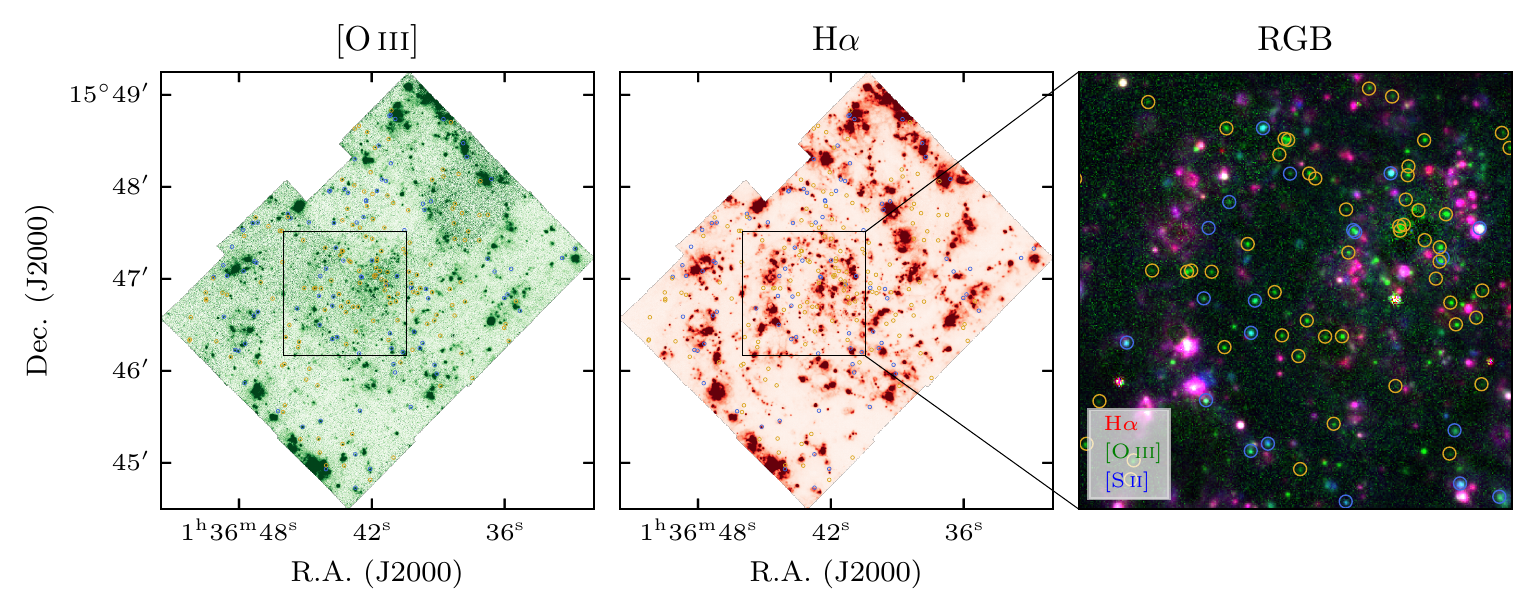}
\caption{Mapping the location of emission lines sources in \galaxyname{NGC}{628}. \pn{}e (marked with gold circles) appear as bright point like objects in the $\OIII$ map (left). Objects that also show significant $\HA$ emission (centre) are often other objects like \HII regions or \snr{}s (marked with blue circles). In the composite RGB image (right), \pn{}e stand out as bright green dots.} 
\label{fig:NGC0628_detections}
\end{figure*}

To create an initial catalogue of \pn{}e candidates, we use a four step procedure: First, we search for unresolved objects in the Gaussian fit $\OIII$ line map. Then, we measure their fluxes within the non-parametric moment 0 maps with aperture photometry and subtract the background from an annulus. Next, we correct for the flux that is lost outside of the aperture and finally apply a correction for Milky Way extinction.

The physical resolution of the observations varies between \SIrange[range-phrase=\ and\ ]{25}{95}{\parsec} \citep[based on prior distances from][]{Anand+2021}. This means that \pn{}e, which typically have sizes of less than $\SI{1}{\parsec}$ \citep{Bojicic+2021}, will always appear as unresolved objects. For the bright \pn{}e that are of interest to us, the strongest emission line is the forbidden $\OIII\,\lambda 5007$ line, so we search for point sources in this line map (see Figure~\ref{fig:NGC0628_detections}). 
Before we start with the source finding, we mask out regions around known stars from \textit{GAIA} DR2 \citep{GaiaCollaboration+2018} to account for foreground objects. Other problematic regions are the centres of galaxies with strong $\OIII$ emission. This can lead to other objects being mistakenly classified as \pn{}e. To avoid this, we create an elliptical mask and place it at the galaxy centre. We select a semi-major-axis of $0.2\cdot r_{25}$. This allows us to reliable get rid of most problematic objects. We also mask the parts of the image that have $\HA$ fluxes above the 95\textsuperscript{th} percentile of the entire image. This removes some parts of the spiral arms. This problem only concerns some of the galaxies in our sample and we only apply this treatment to the following galaxies: \galaxyname{NGC}{1300}, \galaxyname{NGC}{1365}, \galaxyname{NGC}{1512}, \galaxyname{NGC}{1566}, \galaxyname{NGC}{1672}, \galaxyname{NGC}{3351}, \galaxyname{NGC}{3627} \galaxyname{NGC}{4303}, \galaxyname{NGC}{4321}, \galaxyname{NGC}{4535} and \galaxyname{NGC}{7496}. Between \SIrange{6.5}{13.6}{\percent} of the fields of view are masked in the process.

The source detection itself is performed with the Python package \textsc{photutils} \citep{Photutils+2019} which provides an implementation of the \textsc{daophot} algorithm from \citet{Stetson+1987}. Because individual pointings differ slightly in noise and PSF size, we process them individually. For each of them, the noise level is estimated from the standard deviation with iterative three sigma clipping. We then search for objects above a threshold of 3 times this noise level. This constitutes our preliminary catalogue of \pn{} candidates.

The next step is to measure the fluxes, which is done with aperture photometry. A smaller aperture is generally favourable because it only uses the brightest part of the object and hence increases the signal to noise \citep{Howell+1989}. However, this requires a good knowledge of the PSF. Because we are not able to get a precise measurement of the PSF in each pointing \citep[see][ for a detailed discussion of the PSF]{Emsellem+2021}, those errors would then propagate into the measured fluxes. We therefore use a rather large aperture with a diameter of $2.5\cdot\mathrm{FWHM}$. Only with this large aperture size does the uncertainty due to the PSF reach a level similar to that of the photometric error ($\approx \SI{0.1}{\mag}$).

The background is estimated from an annulus, with the inner radius set to $4\cdot\mathrm{FWHM}$ and the outer radius chosen such that the covered area is five times the area of the aperture, corresponding to $6.9\cdot\mathrm{FWHM}$. To remove bright contaminants within the annulus, we apply iterative three sigma clipping. The sigma-clipped median of the annulus is then scaled to the size of the central aperture and subtracted. Afterwards, we correct for flux that is lost outside of the aperture. To do this we integrate the PSF from Equation~\ref{eq:Moffat}. This yields the fraction of flux within an aperture of radius~$r$,
\begin{equation}
    f (r) = 1- \left[ 1+\left(\frac{r^2}{\gamma^2}\right) \right]^{1-\alpha}\;.
\end{equation}
This correction is then applied to all measured fluxes. Compared to the noise in the image, the associated error maps seem to be underestimated. This is confirmed by looking at lines with a fixed theoretical ratio, e.g., $\OIII[4958]/\OIII[5007]=0.35$, $\OI[6363]/\OI[6300]=0.33$ or $\NII[6548]/\NII[6583]=0.34$. By fitting a pair of lines independent from each other and comparing the deviation from the theoretical ratio with the measured uncertainties, we conclude that the uncertainties are underestimated by a factor of $1.38$ to $1.87$ (with larger deviations at shorter wavelengths). To compensate for that, we increase the measured uncertainties by a factor of $1.67$, reflecting the mean correction factor from our tests. The average uncertainty of the $\OIII$ magnitudes after this correction is $\SI{0.078}{\mag}$. To this we add in quadrature the uncertainty that arises from the error of the PSF measurement (estimated to be $\SI{0.1}{\mag}$ on average). This makes up the final value that is reported in Table~\ref{tbl:PN_Identifications} and used throughout the analysis.

Next, all measured fluxes are corrected for Galactic foreground extinction. For the Milky Way foreground we use the \citet{Cardelli+1989} extinction curve with $R_V = 3.1$ and $E(B{-}V)$ from \citet[][see Table~\ref{tbl:sample}]{Schlafly+2011}. 
We choose not to correct for internal extinction. Neither for circumstellar extinction that is associated with the \pn, nor for the dust within the target galaxy. The former is part of the empirical calibration of the luminosity function, and so far only a few studies investigated it \citep{Davis+2018}. For the latter it is common practice not to correct for it, based on a number of arguments \citep{Feldmeier+1997}. In the case of the Milky Way, \pn{}e have a higher vertical scale height \citep{Bobylev+2017} than dust \citep{Li+2018}. If \pn{}e are equally distributed in front of and behind the dust, we will observe about half of the sample with minimal extinction. As long as this subset sufficiently samples the bright end of the \pnlf, we will measure the correct distance. Strictly speaking, this implies that we should observe a different shape, composed of two shifted luminosity functions. In Appendix~\ref{sec:extinction} we show that our algorithm is able to recover the correct distance even when we have a compound luminosity function, consisting of a sample with extinction and one without extinction. On the other hand, even a few bright \pn{}e that are wrongly extinction corrected can significantly reduce the measured distances. 

Finally, the $\OIII$ fluxes (in $\si{\erg \per \s \per \cm \squared}$) are converted to apparent magnitudes with the formula by \citet{Jacoby+1989a}
\begin{equation}
    m_{\OIII} = -2.5 \cdot \log_{10} I_{\OIII} - 13.74\;.  
\end{equation}

\subsection{Comparison with existing studies}\label{sec:detection_comparison}
 
For a handful of galaxies in our sample, the \pnlf{} has been measured in previous studies. This includes \galaxyname{NGC}{0628} \citep{Herrmann+2008,Kreckel+2017,Roth+2021}, \galaxyname{NGC}{3351} and \galaxyname{NGC}{3627} \citep[both][]{Ciardullo+2002} and \galaxyname{NGC}{5068} \citep{Herrmann+2008}. To validate our source detection and photometry, we compare the positions and measured fluxes of our \pn{}e candidates when available. 

For \galaxyname{NGC}{0628}, we recover around $\SI{75}{\percent}$ of the sources (including the 20 brightest objects). In Figure~\ref{fig:NGC0628_fluxes_comparison}, we compare our measured fluxes.  While we find good agreement for the $\OIII$ magnitudes, with only minor offsets of $\Delta m_{\OIII} = \SI{0.03+-0.23}{\mag}$ \citep{Herrmann+2008}, $\Delta m_{\OIII} = \SI{0.01+-0.29}{\mag}$ \citep{Kreckel+2017} and $\Delta m_{\OIII} = \SI{0.08+-0.26}{\mag}$ \citep{Roth+2021}, the $\HA$ fluxes are at odds. Figure~\ref{fig:NGC0628_fluxes_comparison} suggests systematically higher $\HA$ fluxes compared to \citet{Kreckel+2017}. This discrepancy with \citet{Kreckel+2017}, who used the same data as this work, can be attributed to a different technique for the background subtraction. \citet{Kreckel+2017} smoothed the image to $\SI{2}{\arcsec}$ to create a global background map, which is then subtracted from the original image. In our tests, we find that this results in a higher background compared to our annulus background subtraction. The ensuing larger $\HA$ fluxes in our analysis can alter the subsequent classification in favour of \HII regions.

For the other galaxies, the overlap is sparse. Narrowband studies have difficulties in removing the stellar continuum and therefore focus on the outer regions to find \pn{}e while our observations do not extend far beyond the central region. Therefore, only a handful of objects appear in both samples. \galaxyname{NGC}{3351} has six and \galaxyname{NGC}{3627} has one previously detected \pn{}e in our field of view, and we are able to recover them all. Except for the faintest objects, the $\OIII$ fluxes agree within $\SI{0.2}{\mag}$, but no $\HA$ fluxes are reported. For \galaxyname{NGC}{5068}, two of their \pn{}e fall in our field of view but we are only able to detect one of them.

\begin{figure}
\centering
\includegraphics[width=0.9\columnwidth]{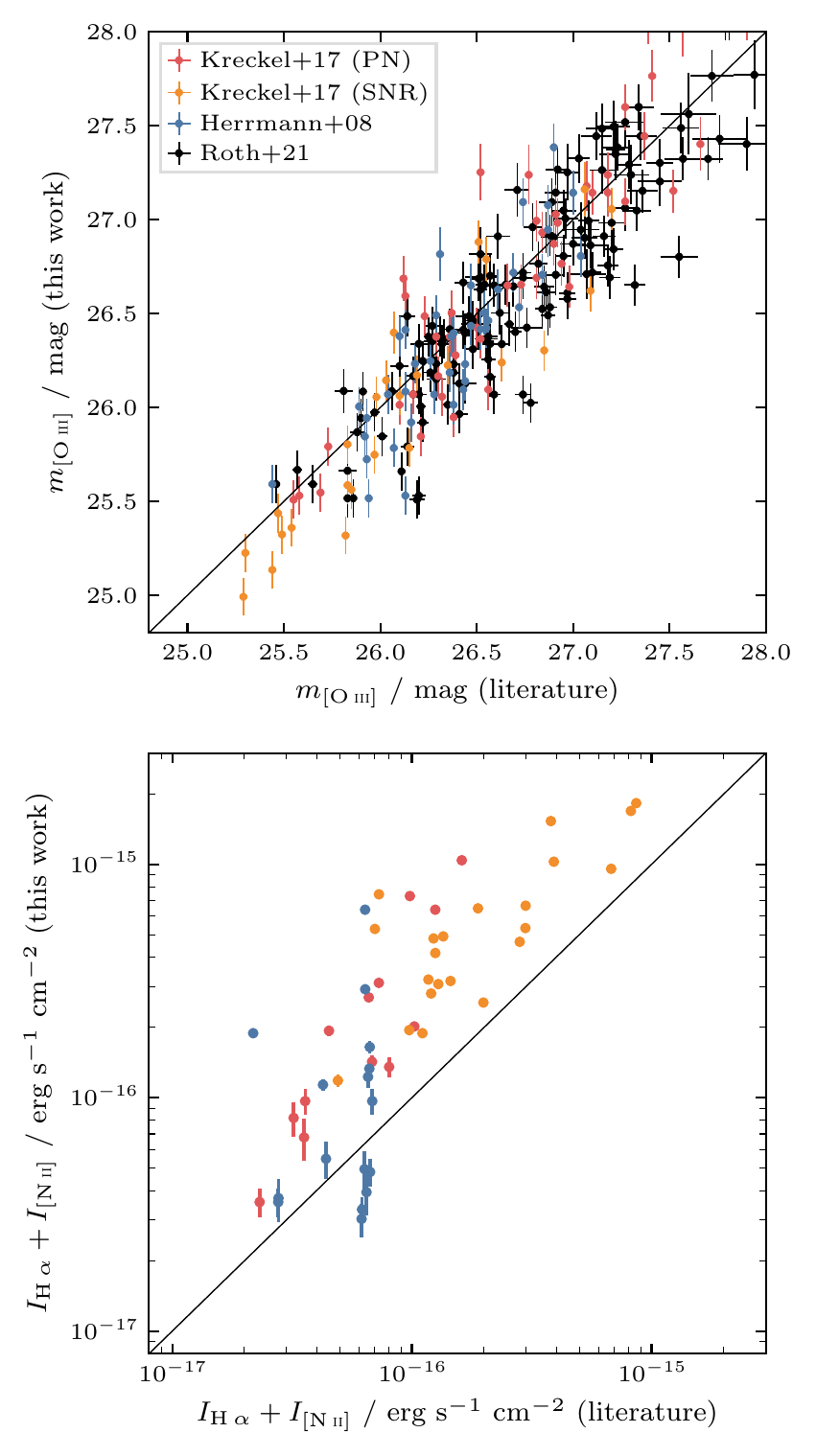}
\caption{Top panel: Comparison of the $\OIII$ apparent magnitudes in \galaxyname{NGC}{0628} between our detections and the catalogues from \citet{Herrmann+2008}, \citet{Kreckel+2017} and \citet{Roth+2021}. The solid line represents equality. Bottom panel: Comparison of the $\HA+\NII$ fluxes in \galaxyname{NGC}{0628}. Narrowband imaging does not resolve the $\HA$ and $\NII$ lines individually and instead measures a wider part of the spectrum. While not identical, we try to replicate this behaviour by summing the $\HA$ and the $\NII$ emission line. The solid line represents equality.}
\label{fig:NGC0628_fluxes_comparison}
\end{figure}

\subsection{Emission line diagnostics}\label{sec:emission_line}

The galaxies in the PHANGS--MUSE sample are, by selection, star-forming. This means that the detected $\OIII$ emission is not exclusively coming from \pn{}e, but can also originate from \HII regions or compact \snr{}s. To eliminate them from our catalogue of \pn candidates, we apply a set of \emph{emission line diagnostics}. The origin of the ionising radiation is different for the three classes of objects (white dwarfs for \pn{}e, O~and B~stars for \HII regions and shocks for \snr{}s). This yields different characteristics in their spectra which can be used to discriminate them. The star at the centre of a bright \pn is much hotter than the star(s) that ionise an \HII region. It has a harder spectrum where more doubly-ionised oxygen is produced compared to $\HA$ \citep{Shaver+1983} and the ratio $\OIII/(\HA+\NII)$ should therefore be greater than~$1.6$ ($\HA$ and $\NII$ are taken together because the lines are close and narrow band imaging can not resolve the individual lines). This effect is even more pronounced for luminous \pn{}e which usually show even higher ratios. Therefore, instead of drawing a straight horizontal line, we use the criterion from \citet{Ciardullo+2002} who showed that \pn{}e typically fall above a line defined by
\begin{equation}\label{eq:HIIregion_criteria}
\log_{10} \frac{I_{\OIII[5007]}}{I_{\HA+\NII\lambda6583}} > -0.37 M_{\OIII} - 1.16 \;.
\end{equation}
This still leaves possible contamination from compact \snr{}s. To eliminate them, we use a criterion from \citet{Blair+2004}. The shock-heated material of \snr{}s will have higher $\SII/\HA$ ratios with
\begin{equation}\label{eq:SNR_criteria}
 \log_{10} \frac{I_{\SII[6717] + \SII[6731]}}{I_{\HA}} > -0.4\;.
\end{equation}
This cut can also remove some \HII regions, but it is unlikely that it removes \pn{}e. The classification based on both criteria is shown in Figure~\ref{fig:NGC628_emission_line_diagnostics}. Many of our \pn candidates are detected neither in $\HA$ nor in $\SII$ (we call everything below $3\sigma$ undetected). Objects that are consistent within the uncertainty of being a \pn are retained in the sample.

A last criterion is given by the velocity dispersion of the emission lines. The shell of a \pn expands slowly (${\sim}\SI{20}{\km\per\second}$) into the ambient medium \citep{Schoenberner+2005} whereas \snr{}s have much higher expansion velocities \citep[${\sim}\SI{100}{\km\per\second}$,][]{Franchetti+2012}. We measure the velocity dispersion from the $\OIII$ line, averaged over the central pixels of the object, and deconvolve it from the instrumental resolution. We note that, while many objects have quite high signal to noise, the instrumental spectral resolution is much larger (${\sim}\SI{50}{\km \per \second}$) than the intrinsic dispersion and so our deconvolution is sensitive to the shape of the assumed line spread function.  Because of this, we are confident in relative velocity dispersions, but have reservations about their absolute values. We therefore decide not to use this criterion for the classification, but only as an additional check. Indeed, \snr{}s ($\SI{61}{\km \per \second}$) have a larger velocity dispersion than \pn{}e ($\SI{48}{\km \per \second}$) or \HII regions ($\SI{30}{\km \per \second}$).

\begin{figure}
\centering
\includegraphics[width=0.9\columnwidth]{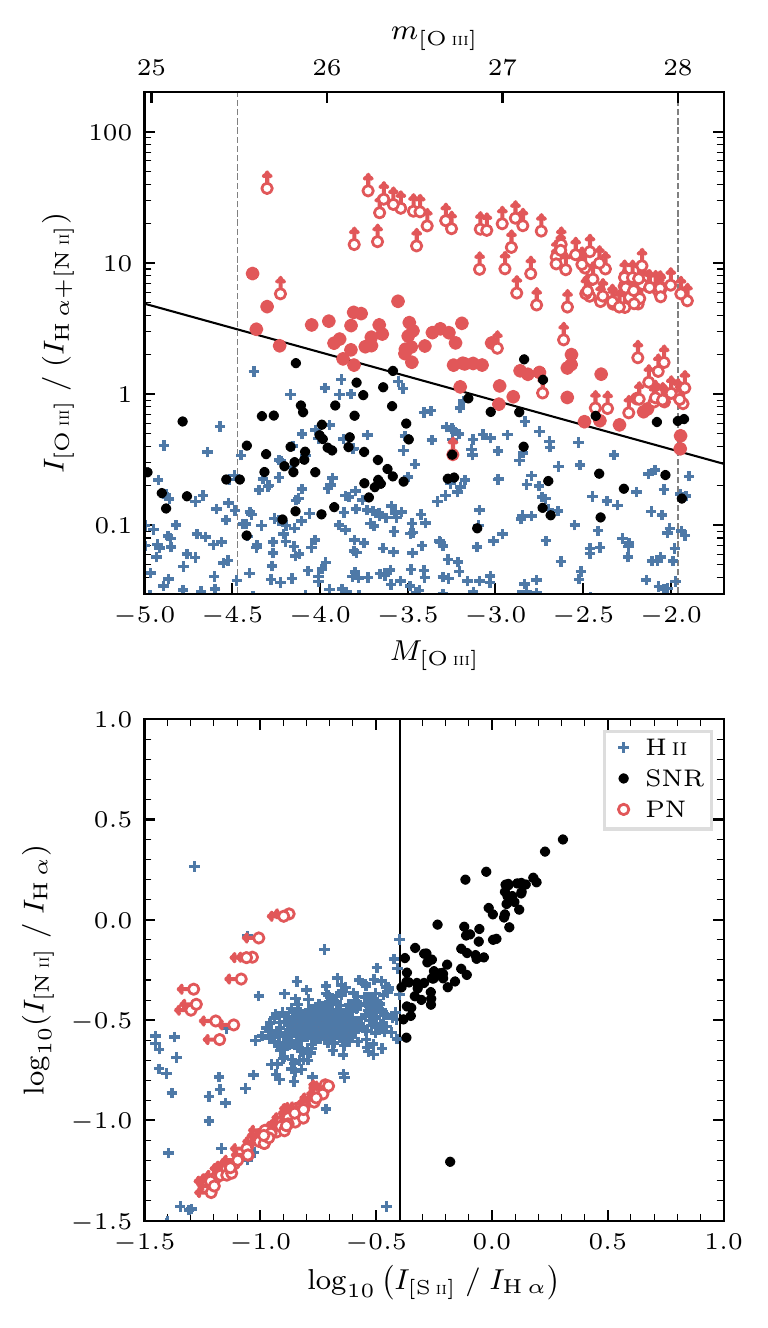}
\caption{Emission line diagnostics for \galaxyname{NGC}{0628}. The top panel shows the criterion from Equation~\ref{eq:HIIregion_criteria} that is used to eliminate \HII regions from our sample. Open circles indicate when a \pn{}e is not detected in $\HA$ and the symbol is a lower limit. The lower panel shows the criterion from Equation~\ref{eq:SNR_criteria} that is used to remove \snr{}s. Open circles indicate that a \pn{}e is not detected in $\SII$ and the symbol is an upper limit.}
\label{fig:NGC628_emission_line_diagnostics}
\end{figure}

\begin{table*}
\centering
\caption{Planetary nebula identifications and supernova remnant contaminants. The full table is available in the online supplementary material.}
\sisetup{
parse-numbers = false
}
\begin{tabular}{
            l 
            r
            r
            r
            r
            S[table-format=+2.2]@{\,\( \pm \)\,}
            S[table-format=1.2]
            S[table-format=+1.2]@{\,\( \pm \)}
            S[table-format=1.2]
            S[table-format=+1.2]@{\,\( \pm \)}
            S[table-format=1.2]
            S[table-format=+1.2]@{\,\( \pm \)}
            S[table-format=1.2]
            c
            }
\toprule\toprule
Galaxy & ID & Type &  \multicolumn{1}{c}{R.A.} & \multicolumn{1}{c}{Dec.} & \multicolumn{2}{c}{$m_{\OIII}$} & \multicolumn{2}{c}{$\log (I_{\OIII}/I_{\HA})$} & \multicolumn{2}{c}{$\log (I_{\NII}/I_{\HA})$} & \multicolumn{2}{c}{$\log ( I_{\SII}/I_{\HA})$} & Notes \\
& & & \multicolumn{1}{c}{(J2000)} & \multicolumn{1}{c}{(J2000)} & \multicolumn{2}{c}{mag} & \multicolumn{2}{c}{ } & \multicolumn{2}{c}{ } & \multicolumn{2}{c}{ } &  \\
\midrule
\galaxyname{NGC}{0628} & 1 & PN & 01h36m43.31s & +15d47m18.15s & 25.51 & 0.1 & 1.1 & 0.07 & -0.83 & 0.44 & -0.71 & 0.44 &  \\
\galaxyname{NGC}{0628} & 2 & PN & 01h36m43.35s & +15d47m18.35s & 25.53 & 0.1 & 0.67 & 0.03 & -1.24 & 0.44 & -1.13 & 0.44 &  \\
\galaxyname{NGC}{0628} & 3 & PN & 01h36m41.29s & +15d47m04.41s & 25.59 & 0.1 & 1.75 & 0.43 & -0.03 & 0.61 & 0.05 & 0.61 &  \\
\galaxyname{NGC}{0628} & 4 & PN & 01h36m39.96s & +15d47m02.64s & 25.59 & 0.1 & 0.84 & 0.04 & -1.14 & 0.44 & -1.06 & 0.44 &  \\
\galaxyname{NGC}{0628} & 5 & PN & 01h36m41.37s & +15d46m58.31s & 25.66 & 0.1 & 0.54 & 0.03 & -1.24 & 0.44 & -1.15 & 0.44 &  \\
\galaxyname{NGC}{0628} & 6 & PN & 01h36m41.37s & +15d46m55.67s & 25.67 & 0.1 & 1.71 & 0.43 & 0.9 & 0.44 & 0.06 & 0.61 &  \\
\galaxyname{NGC}{0628} & 7 & SNR & 01h36m42.48s & +15d47m01.48s & 25.76 & 0.1 & 1.83 & 0.43 & 1.58 & 0.43 & 1.35 & 0.43 &  \\
\galaxyname{NGC}{0628} & 8 & PN & 01h36m43.35s & +15d48m04.68s & 25.85 & 0.1 & 0.7 & 0.04 & -1.12 & 0.44 & -0.98 & 0.44 &  \\
\galaxyname{NGC}{0628} & 9 & PN & 01h36m35.82s & +15d46m19.60s & 25.94 & 0.1 & 0.65 & 0.03 & -0.6 & 0.09 & -1.18 & 0.44 &  \\
\galaxyname{NGC}{0628} & 10 & PN & 01h36m37.95s & +15d46m11.46s & 25.97 & 0.1 & 0.49 & 0.02 & -0.58 & 0.06 & -0.53 & 0.07 &  \\
& $\vdots$ &  & $\vdots$ & $\vdots$ & \multicolumn{2}{c}{$\vdots$} & \multicolumn{2}{c}{$\vdots$} & \multicolumn{2}{c}{$\vdots$} & \multicolumn{2}{c}{$\vdots$} & $\vdots$ \\ 
\galaxyname{NGC}{0628}  & 145 & PN & 01h36m38.39s & +15d48m24.20s & 27.95 & 0.2 & -0.24 & 0.08 & -1.27 & 0.44 & -1.14 & 0.44 &  \\
\galaxyname{NGC}{0628}  & 146 & PN & 01h36m36.17s & +15d47m06.59s & 27.95 & 0.19 & -0.14 & 0.07 & -1.2 & 0.44 & -1.1 & 0.44 &  \\
\galaxyname{NGC}{0628}  & 147 & PN & 01h36m42.34s & +15d47m56.95s & 27.95 & 0.21 & 0.94 & 0.44 & -0.06 & 0.61 & 0.08 & 0.61 &  \\
\galaxyname{NGC}{0628}  & 148 & SNR & 01h36m40.53s & +15d47m31.74s & 27.97 & 0.17 & 0.14 & 0.07 & 0.06 & 0.06 & -0.01 & 0.07 &  \\
\galaxyname{NGC}{0628}  & 149 & PN & 01h36m42.53s & +15d47m33.41s & 27.99 & 0.21 & 0.89 & 0.44 & -0.05 & 0.61 & 0.06 & 0.61 &  \\
\bottomrule
\multicolumn{9}{l}{OL: rejected as overluminous sources.} \\
\end{tabular}
\label{tbl:PN_Identifications}
\end{table*}

\subsection{Overluminous sources}

The resulting \pn catalogue is then visually inspected to exclude any non-point-like objects that slipped through the previous selection process. Most objects that are rejected are removed because their azimuthally averaged radial profil (average flux in annuli of increasing radii, from now on referred to as radial profile) exposes them to be not point-like or they fall very close to another bright object, hampering our ability to measure the flux, but we also have to remove a few objects purely based on their luminosity. 

These `\emph{overluminous}' objects are distinctly brighter than the bright end of the remaining luminosity function (see Section~\ref{sec:results}, Figure~\ref{fig:all_galaxies_PNLF}), and hence, cannot be described by the \pnlf. They have been found in many \pn studies \citep{Longobardi+2013,Hartke+2017,Roth+2021}, but their exact origin remains unclear \citep{Jacoby+1996}. Possible explanations include two \pn{}e that accidentally fall close to each other, a misclassified \HII region or a background \mbox{Ly-$\alpha$} galaxy. If such an object falls close to the bright end of the luminosity function, there is no clear way to eliminate it from the sample and it can lead us to underestimate the distances. However, if a gap exists to the remaining sample, it is easy to exclude such an object because the quality of the resulting fit would be very poor. In total, eight objects are rejected as being overluminous. RGB composites of all eight objects are shown in Figure~\ref{fig:all_galaxies_overluminous}. Three of them are \pn{}e and, therefore, also appear in Figure~\ref{fig:all_galaxies_PNLF}, while the other five are \snr{}s that could be classified as \pn{}e.

\subsection{Final \pn sample}

Based on these detection and selection criteria, we classify $899$~objects as \pn{}e across the 19 PHANGS--MUSE galaxies. A further $150$~objects are classified as \snr{}s, which could contaminate narrowband PNe studies, i.e.~they satisfy the criterion from Equation~\ref{eq:HIIregion_criteria} but not the one from Equation~\ref{eq:SNR_criteria}. The final catalogue is presented in Table~\ref{tbl:PN_Identifications}. The table contains the \pn{}e that are brighter than the completeness limit, and are used in our fit to the \pnlf. The position is reported, together with the apparent $\OIII$ magnitude and the line ratios used to make the classification.  We also tabulate the \snr{}s that would be classified as \pn{}e without the $\SII$ criterion. Special objects like overluminous \pn{}e are flagged with ‘OL’.

\begin{figure*}
\centering
\includegraphics[width=\textwidth]{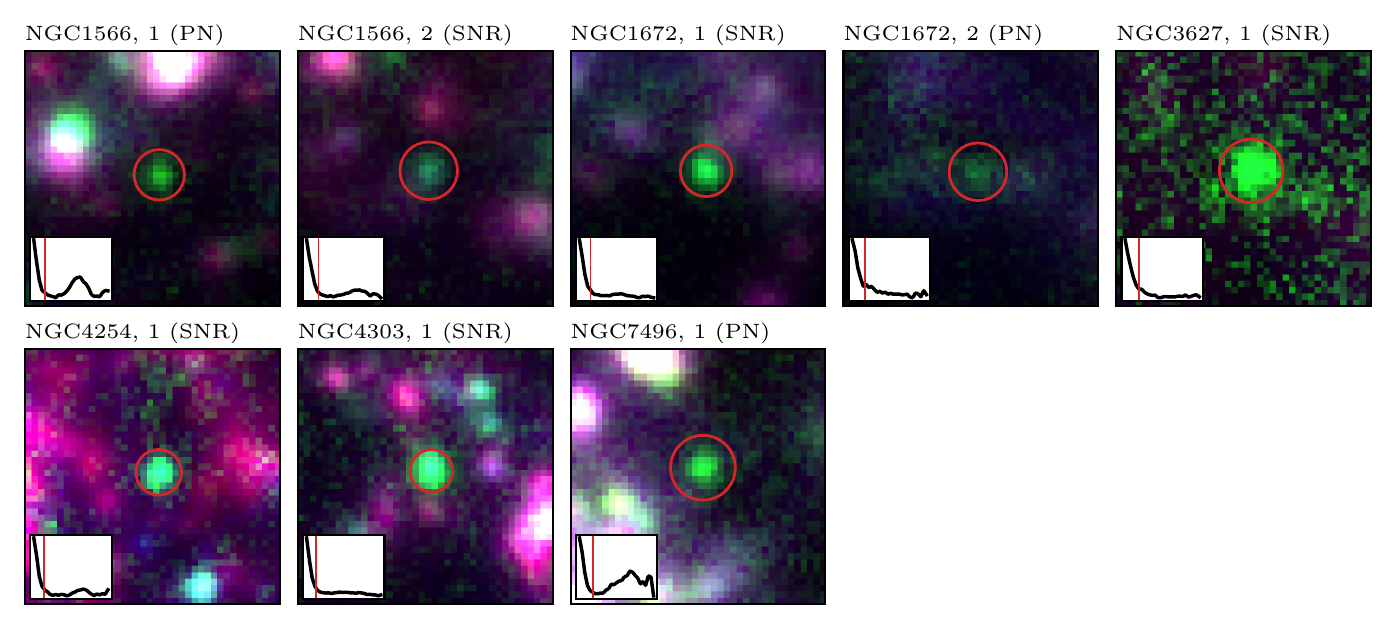}
\caption{In total eight objects were marked and rejected as overluminous objects. Three of them are \pn{}e and five are \snr{}s. The cutouts show composite RGB images of those objects with $\HA$ in red, $\OIII$ in green and $\SII$ in blue. The bottom left corner shows the azimuthally averaged radial profile measured from the $\OIII$ line map. The title contains the name of the galaxy, the `id' in the catalogue (Table~\ref{tbl:PN_Identifications}) and the type of the object.} 
\label{fig:all_galaxies_overluminous}
\end{figure*}

\section{The planetary nebula luminosity function}
\label{sec:pnlf}

The \emph{planetary nebula luminosity function} is an empirical relation that describes the number of \pn{}e that we expect to observe at a certain luminosity. \citet{Ciardullo+1989a} combined an exponential function from theoretical evolutionary models with a cutoff at the bright end for the following formula:
\begin{equation}
    N (M_{\OIII}) \propto \mathrm{e}^{0.307\, M_{\OIII}} \left(1 - \mathrm{e}^{3(M^{*}- M_{\OIII})}\right)~,
	\label{eq:pnlf}
\end{equation}
where $M^*$ is the zero point of the luminosity function. This simple form of the luminosity function has prevailed even though other parametrisations have been proposed. \citet{Hartke+2017}, for example, treated the slope of the faint end ($0.307$ in the original model) as a free parameter and it is even possible to construct a completely numerical luminosity function \citep[e.g.,][]{Mendez+2001,Teodorescu+2011}.

The zero point of the luminosity function, $M^*$, determined by the luminosity of the brightest \pn, has more weight in determining the distance than the functional form itself. This value has to be calibrated from \pn{}e in galaxies with known distances. The original value of $M^*=\SI{-4.48}{\mag}$ was derived based on the Cepheid distance to \galaxyname{M}{31} by \citet{Ciardullo+1989a}. A later study by \citet{Ciardullo+2002} used galaxies with distances from the \textit{HST} Key Project \citep{Freedman+2001} to quantify the impact of metallicity on the zero point. They found a modest dependence with an increase at lower metallicities. The metallicity of the galaxies in our sample varies from $12+\log (\mathrm{O}/\mathrm{H}) = 8.41\ \text{to}\ 8.63$ in the centre and  $12+\log (\mathrm{O}/\mathrm{H}) = 8.00\ \text{to}\ 8.62$ at $r_{25}$ \citep{Kreckel+2019,Santoro+2021}. Since the expected change in $M^*$ is rather small, we decide to adopt a constant value of $M^*=\SI{-4.47}{\mag}$ as a starting point. In Section~\ref{sec:zeropoint}, we use our measured distances and compare them to existing distances from the tip of the red giant branch (\trgb) to review this relation.

\subsection{Completeness limit}

\begin{figure}
\centering
\includegraphics[width=\columnwidth]{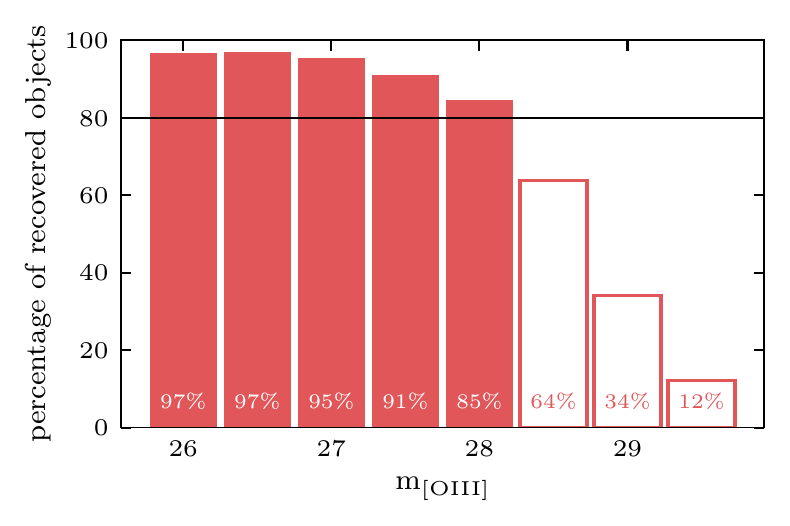}
\caption{Determining the completeness limit for \galaxyname{NGC}{0628}. We assume our sample to be complete when at least $\SI{80}{\percent}$ of the sources are recovered ($m_{\OIII}=\SI{28}{\mag}$ in this case).}
\label{fig:NGC628_completeness}
\end{figure}

Before we fit our data to the luminosity function, we need to estimate the completeness of our sample. Due to noise in the $\OIII$ image and blending with the background, we will miss some of the fainter objects and hence, our luminosity function will not be fully sampled at lower luminosities. To estimate the luminosity where this becomes problematic, we inject mock sources, sampled from a theoretical luminosity function, into the $\OIII$ map and record how many we are able to recover. We never recover $\SI{100}{\percent}$ because the mock sources are randomly placed in the image and sometimes fall onto crowded areas where they are not detectable. Figure~\ref{fig:NGC628_completeness} shows the fraction of recovered objects as a function of their $\OIII$ magnitude. 

One could use this function and `multiply' it with the luminosity function to account for the sources that were missed. Instead, we use an easier approach and fit only the subset of objects that are brighter than a certain threshold. We find that for most galaxies we are able to recover at least $\SI{80}{\percent}$ of the objects brighter than a limiting magnitude of $\SI{28}{\mag}$. Including or excluding a few objects at the faint end does not alter the measured distance and we therefore adopt a uniform completeness limit of $m_{\OIII}=\SI{28}{\mag}$. There are two exceptions, \galaxyname{NGC}{2835} and \galaxyname{NGC}{3627}. For these two galaxies, the $\SI{80}{\percent}$ completeness required by us is only recovered for magnitudes brighter than $m_{\OIII}=\SI{27.5}{\mag}$.

\subsection{Fitting the data}

We can write the absolute magnitude in terms of the observed apparent magnitude and the distance modulus $\mu=(m-M)$ and fit Equation~\ref{eq:pnlf} to our observed data. We normalise the function to the number of observed \pn{}e from the root ($=M^*+\mu$) up to the previously determined completeness limit.  The fitting itself is done via the method of \emph{maximum likelihood estimation} \citep[MLE; see e.g. ][for a detailed introduction]{Hogg+2010}. This alleviates the uncertainties that are introduced when binning the data for a least $\chi^2$ fit. For the MLE, we define the likelihood of a parameter, given some data, as the product of the individual probabilities (or in this case the sum of the log likelihood)
\begin{equation}
    \mathcal{L} (\mu) = \sum_i \log p(m_i|\mu)~.
\end{equation}
The maximum of this function yields the parameters that best describe our observed data. To account for uncertainties in the observed magnitudes, we calculate the probability of observing a \pn{}e of apparent magnitude $m_{\OIII}$ as the integral of the luminosity function multiplied by a Gaussian centred around the magnitude of the \pn with the uncertainty of the apparent magnitude $\sigma_m$ as the width
\begin{equation}
    p(m_{\OIII}) = \int N(m') \frac{1}{\sqrt{2\pi}\sigma_m} \exp \left[ {-\frac{(m'-m_{\OIII})^2}{2\sigma_m^2}}  \right] \mathrm{d} m'~.
	\label{eq:pnlf_with_error}
\end{equation}
Strictly speaking, the uncertainties of the magnitudes are not Gaussian (assuming that the uncertainties of the fluxes are). However for sufficiently high signal to noise ($>10$), the distribution approaches a normal distribution, which is the case for most of our objects. The effect of this procedure is only marginal, but it can take away some weight from the brightest \pn{}e that usually dominate the outcome of the fit. The uncertainty of the fit is taken from the $\SI{68}{\percent}$ confidence interval of the likelihood function. In Appendix~\ref{sec:pnlf_stat} we look in more detail at the fitting procedure and the reliability of the \pnlf.

The classification criterion in Equation~\ref{eq:HIIregion_criteria} depends on a prior estimate of the distance, for which we use the distances from Table~\ref{tbl:sample} \citep{Anand+2021} as a starting point. If our measured distance deviates significantly from this value, the classification of some objects might change. To account for this, we use an iterative process where we classify and fit repeatedly until the classification and the measured distance does not change anymore. 

\section{Results}
\label{sec:results}

In this section, we present an overview of the measured distances and discuss the quality of the individual fits. For each galaxy, we first fit the \pnlf to the clean \pn{}e sample and adopt this value as our preferred distance. We then also measure the distance from the catalogue with the potential \snr contaminants to quantify how misclassified \snr{}s would impact the result. Both values are presented in Table~\ref{tbl:distances}. Figure~\ref{fig:all_galaxies_PNLF} shows the \pnlf for all galaxies in our sample and Figure~\ref{fig:all_galaxies_PNLF_cum} shows their cumulative luminosity function. We use a \emph{Kolmogorov--Smirnov test} to compare the observed sample to a theoretical luminosity function. A high test statistic, $D_\mathrm{max}$ (the maximum distance between the empirical distribution function and the cumulative distribution function of the underlying model), and a low $p$-value rejects the null hypothesis that the observed data were drawn from the theoretical distribution. The quality of the \pnlf{}s varies greatly in our sample. While the closer galaxies are well sampled, the quality of the fit deteriorates with distance. Beyond $\SI{15}{\mega\parsec}$, it becomes increasingly difficult to detect a sufficient number of \pn{}e. This corresponds to a distance modulus of $(m-M)=\SI{30.9}{\mag}$, which puts the bright end cutoff at $m_{\OIII}=\SI{26.4}{\mag}$. With a completeness limit of $\SI{28}{\mag}$, this means we are only able to observe the upper $\SI{1.6}{\mag}$ of the luminosity function. The reported values for those galaxies should therefore be treated with caution. Figure~\ref{fig:NGC0628_distances} shows a comparison between our measured distances with a compilation of literature distance for \galaxyname{NGC}{0628}. Similar plots for the full sample are shown in Appendix~\ref{sec:literature_distances}.\medskip

\begin{table}
    \centering
    \caption{Measured \pnlf distances. $N_\text{\pn}$ is the number of detected \pn{}e brighter than our completeness limit and $N_\text{\snr}$ is the number of \snr{}s that could be misclassified as \pn{}e. $(m-M)$ and $(m-M)_{\snr}$ are the distance moduli measured without and with \snr contamination, respectively. 
    }
{\renewcommand{\arraystretch}{1.4} 
    \begin{tabular}{lccrrr}\toprule\toprule
Name & $N_\text{\pn}$ & $N_\text{\snr}$ & \multicolumn{1}{c}{$(m-M)$} & \multicolumn{1}{c}{$(m-M)_\text{SNR}$} & \multicolumn{1}{c}{Distance}  \\
& & & \multicolumn{1}{c}{$\si{mag}$} & \multicolumn{1}{c}{$\si{mag}$} & \multicolumn{1}{c}{$\si{\mega\parsec}$}  \\\midrule
\galaxyname{IC}{5332} & 47 & 16 & $\uncertainty{29.73}{0.10}{0.20}$ & $\uncertainty{29.78}{0.09}{0.16}$ & $\uncertainty{8.84}{0.39}{0.82}$ \\
\galaxyname{NGC}{0628} & 139 & 10 & $\uncertainty{29.89}{0.06}{0.09}$ & $\uncertainty{29.90}{0.06}{0.09}$ & $\uncertainty{9.52}{0.26}{0.41}$ \\
\galaxyname{NGC}{1087} & 15 & 5 & $\uncertainty{31.05}{0.10}{0.24}$ & $\uncertainty{31.06}{0.09}{0.20}$ & $\uncertainty{16.25}{0.74}{1.79}$ \\
\galaxyname{NGC}{1300} & 17 & 3 & $\uncertainty{32.06}{0.08}{0.12}$ & $\uncertainty{32.03}{0.07}{0.11}$ & $\uncertainty{25.77}{0.90}{1.42}$ \\
\galaxyname{NGC}{1365} & 29 & 5 & $\uncertainty{31.22}{0.08}{0.14}$ & $\uncertainty{31.19}{0.07}{0.13}$ & $\uncertainty{17.53}{0.66}{1.16}$ \\
\galaxyname{NGC}{1385} & 11 & 9 & $\uncertainty{29.96}{0.14}{0.32}$ & $\uncertainty{30.04}{0.12}{0.28}$ & $\uncertainty{9.81}{0.63}{1.46}$ \\
\galaxyname{NGC}{1433} & 90 & 6 & $\uncertainty{31.39}{0.04}{0.07}$ & $\uncertainty{31.39}{0.04}{0.06}$ & $\uncertainty{18.94}{0.39}{0.56}$ \\
\galaxyname{NGC}{1512} & 43 & 5 & $\uncertainty{31.27}{0.07}{0.11}$ & $\uncertainty{31.29}{0.06}{0.10}$ & $\uncertainty{17.93}{0.53}{0.88}$ \\
\galaxyname{NGC}{1566} & 27 & 2 & $\uncertainty{31.13}{0.08}{0.17}$ & $\uncertainty{31.14}{0.08}{0.16}$ & $\uncertainty{16.84}{0.60}{1.29}$ \\
\galaxyname{NGC}{1672} & 19 & 2 & $\uncertainty{30.99}{0.09}{0.23}$ & $\uncertainty{30.98}{0.09}{0.22}$ & $\uncertainty{15.80}{0.68}{1.68}$ \\
\galaxyname{NGC}{2835} & 27 & 1 & $\uncertainty{30.57}{0.08}{0.17}$ & $\uncertainty{30.57}{0.08}{0.17}$ & $\uncertainty{13.03}{0.46}{1.04}$ \\
\galaxyname{NGC}{3351} & 142 & 10 & $\uncertainty{30.36}{0.06}{0.08}$ & $\uncertainty{30.32}{0.06}{0.08}$ & $\uncertainty{11.80}{0.31}{0.43}$ \\
\galaxyname{NGC}{3627} & 43 & 7 & $\uncertainty{30.18}{0.08}{0.15}$ & $\uncertainty{30.16}{0.07}{0.14}$ & $\uncertainty{10.88}{0.39}{0.77}$ \\
\galaxyname{NGC}{4254} & 42 & 22 & $\uncertainty{29.97}{0.09}{0.20}$ & $\uncertainty{29.91}{0.08}{0.16}$ & $\uncertainty{9.86}{0.42}{0.91}$ \\
\galaxyname{NGC}{4303} & 19 & 7 & $\uncertainty{30.65}{0.10}{0.26}$ & $\uncertainty{30.17}{0.10}{0.25}$ & $\uncertainty{13.49}{0.64}{1.60}$ \\
\galaxyname{NGC}{4321} & 62 & 11 & $\uncertainty{31.10}{0.06}{0.10}$ & $\uncertainty{30.78}{0.08}{0.11}$ & $\uncertainty{16.62}{0.46}{0.74}$ \\
\galaxyname{NGC}{4535} & 53 & 0 & $\uncertainty{31.43}{0.06}{0.09}$ & $\uncertainty{31.43}{0.06}{0.09}$ & $\uncertainty{19.29}{0.56}{0.82}$ \\
\galaxyname{NGC}{5068} & 58 & 22 & $\uncertainty{28.46}{0.11}{0.26}$ & $\uncertainty{28.54}{0.09}{0.21}$ & $\uncertainty{4.93}{0.24}{0.59}$ \\
\galaxyname{NGC}{7496} & 13 & 2 & $\uncertainty{31.64}{0.09}{0.19}$ & $\uncertainty{31.04}{0.11}{0.23}$ & $\uncertainty{21.31}{0.89}{1.89}$ \\
\bottomrule
\end{tabular}
}
\label{tbl:distances}
\end{table}

\begin{figure*}
\centering
\includegraphics[width=0.99\textwidth]{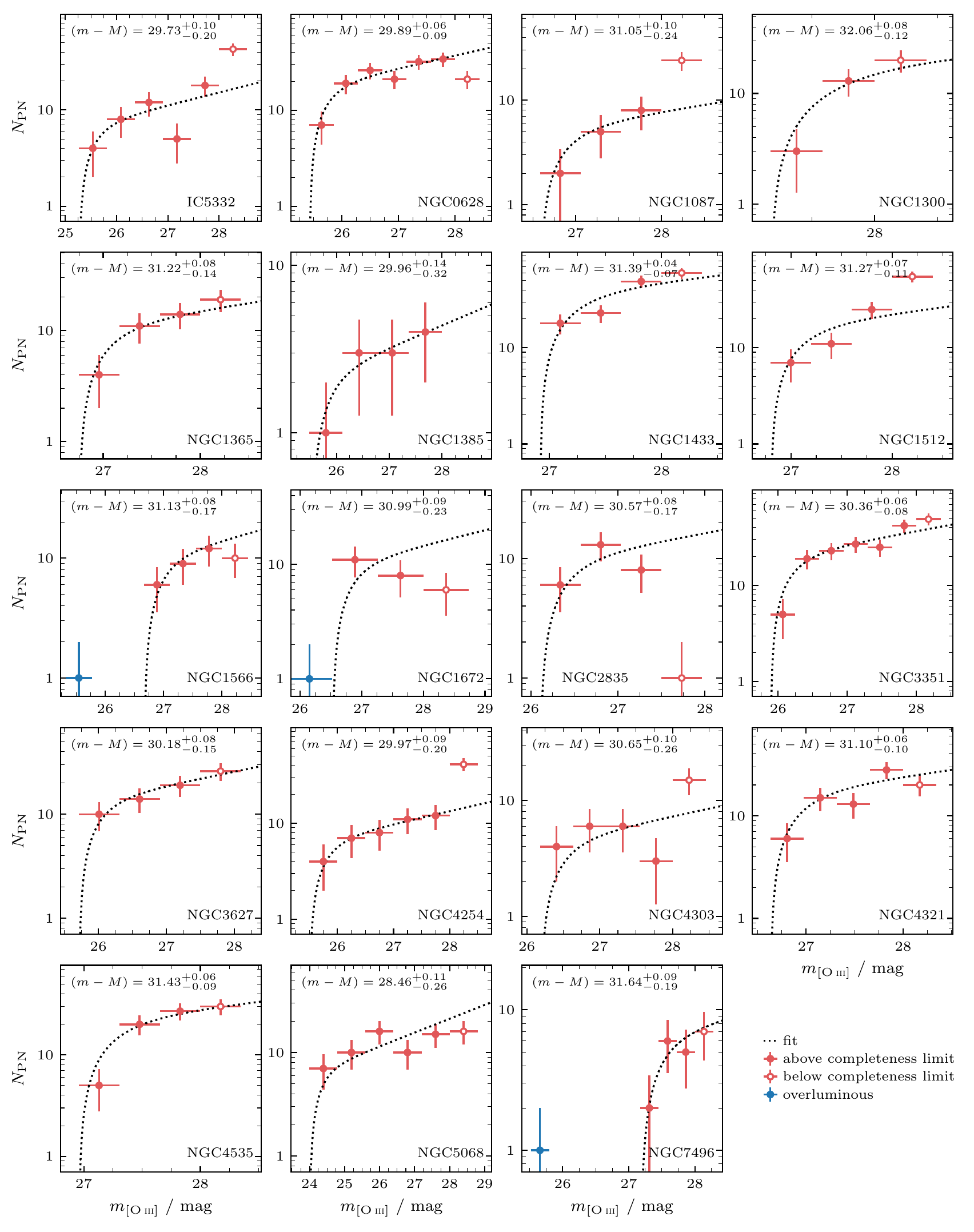}
\caption{\pnlf for all 19 galaxies in our sample. The filled circles denote objects brighter than our completeness limit that contribute to the fit. The $y$-errors are from Poisson statistics and the $x$-errors are the bin widths. The fit of Equation~\ref{eq:pnlf} is done via the method of maximum likelihood. This binning is only used to illustrate the result and not used in the fit. In blue are objects that were rejected as overluminous objects.} 
\label{fig:all_galaxies_PNLF}
\end{figure*}

\begin{figure*}
\centering
\includegraphics[width=\textwidth]{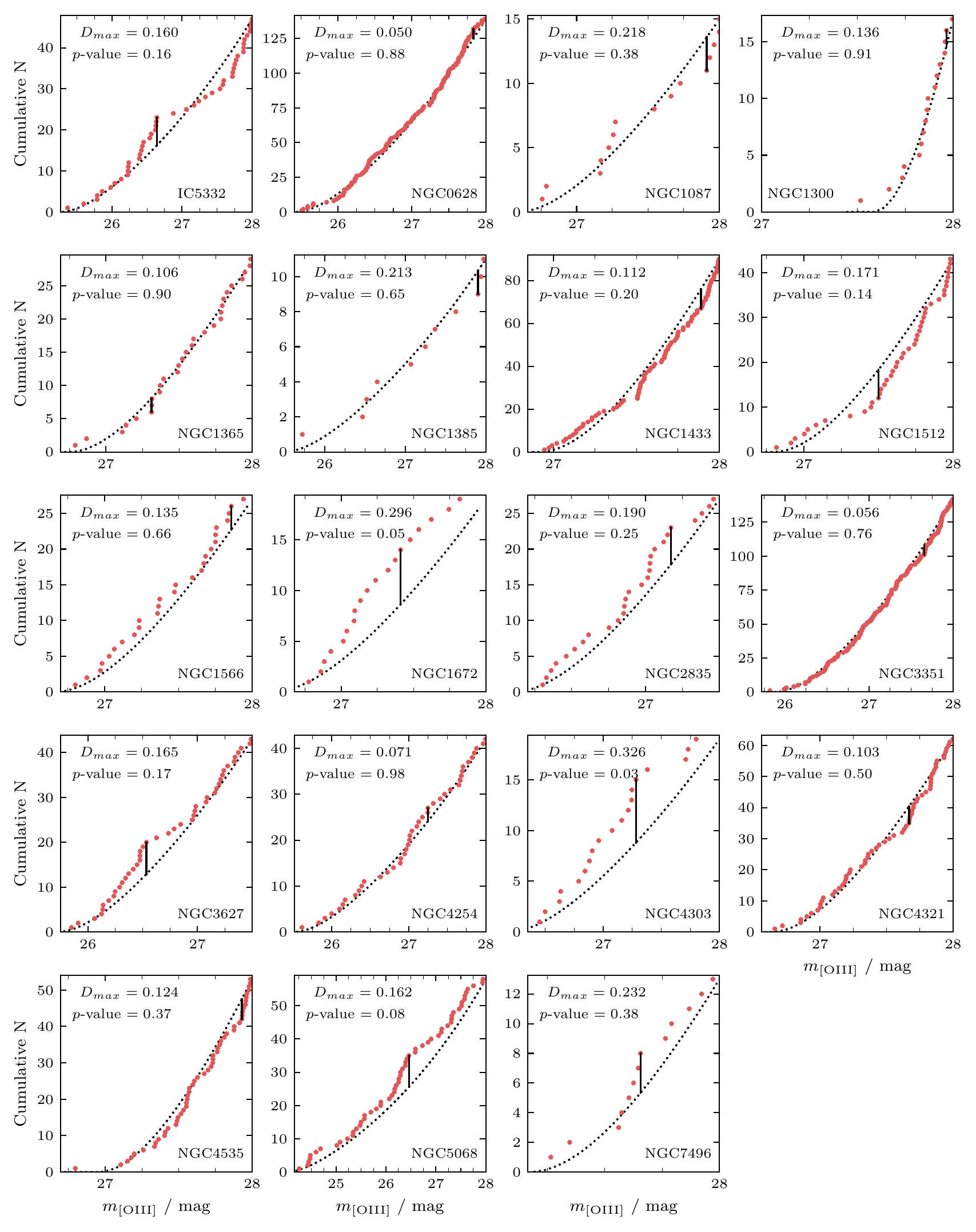}
\caption{Cumulative PNLF for all 19 galaxies in our sample. In the top left corner is the statistic $D_\mathrm{max}$ and the $p$-value from the KS test. The former is indicated with a vertical black bar in the plot (not normalised). The fitted luminosity function is shown with a black dotted line.} 
\label{fig:all_galaxies_PNLF_cum}
\end{figure*}

\begin{figure}
\centering
\input{fig/NGC0628_distances.pgf}
\caption[Caption for LOF (not used but otherwise I get an error with the footnote)]{Comparison between our measured distance and literature values for \galaxyname{NGC}{0628} (\galaxyname{M}{74}). The value measured by this study is marked by a black line with the one and three sigma intervals shaded in grey. We measure a distance modulus of $(m-M)=\SI[parse-numbers=false]{\getvalue{mag/NGC0628}}{\mag}$ ($\SI[parse-numbers=false]{\getvalue{Mpc/NGC0628}}{\mega\parsec}$). A list of abbreviations and similar figures for the other galaxies can be found in Appendix~\ref{sec:literature_distances}.}
\label{fig:NGC0628_distances}
\end{figure}

\textbf{\galaxyname{IC}{5332}} is located in the constellation Sculptor. Until recently, the only available distances came from the Tully--Fisher method and they show a large scatter, with the newest reported distance modulus of $\SI{29.62+-0.4}{\mag}$ \citep{1988NBGC.C....0000T} being significantly larger than the old ones (see Figure~\ref{fig:IC5332_distances}). A new \trgb distance from the PHANGS--\textit{HST} program \citep{Anand+2021} finds a distance modulus of $\SI{29.77 \pm 0.09}{\mag}$. This places this galaxy among the closest ones in our sample. Our sample consists of $49$~\pn{}e that are brighter than our completeness limit. Two objects were excluded because of their irregular radial profile, but neither of them fall at the bright end, and hence, this does not impact the measured distance. From the remaining $47$~\pn{}e we measure a distance modulus of $(m-M)=\SI[parse-numbers=false]{\getvalue{mag/IC5332}}{\mag}$ ($D=\SI[parse-numbers=false]{\getvalue{Mpc/IC5332}}{\mega\parsec}$). This is in excellent agreement with the previously mentioned \trgb value and the latest Tully--Fisher distance. There are $16$~\snr{}s that would be classified as \pn{}e without the $\SII$ line ratio, but none of them fall at the bright end of the luminosity function and their inclusion does not alter the measured distance significantly. \medskip

\textbf{\galaxyname{NGC}{0628}}, also known as \galaxyname{M}{74}, is located in the constellation Pisces. This galaxy has been studied extensively with numerous existing distance measurements from different methods (see Figure~\ref{fig:NGC0628_distances} for a selection). Among them are the aforementioned \pnlf studies by \citet{Herrmann+2008}, \citet{Kreckel+2017} and \citet{Roth+2021}. This makes this galaxy a valuable benchmark to validate our methods. The distance of \citet{Kreckel+2017} is based on the same PHANGS--MUSE data, but was limited to the three pointings that were available at the time, and \citet{Roth+2021} uses the same raw data but with a different data reduction. We find $139$~\pn{}e that provide a well constrained distance modulus of $(m-M)=\SI[parse-numbers=false]{\getvalue{mag/NGC0628}}{\mag}$ ($D=\SI[parse-numbers=false]{\getvalue{Mpc/NGC0628}}{\mega\parsec}$). This is in excellent agreement to the result from \citet{Kreckel+2017} of $(m-M)=\SI[parse-numbers=false]{\uncertainty{29.91}{0.08}{0.13}}{\mag}$, while both \citet{Herrmann+2008} and \citet{Roth+2021} derived a smaller distance modulus of $(m-M)=\SI[parse-numbers=false]{\uncertainty{29.67}{0.06}{0.07}}{\mag}$ and $(m-M)=\SI[parse-numbers=false]{\uncertainty{29.76}{0.03}{0.05}}{\mag}$, respectively. The discrepancy was explained by \citet{Kreckel+2017} with \snr{}s that were misclassified as \pn{}e. We are unable to confirm this explanation. Because we measure larger $\HA$ fluxes (see also Section~\ref{sec:detection_comparison}), the objects in question are mostly classified as \HII regions in our analysis (see Figure~\ref{fig:NGC628_emission_line_diagnostics}) and, therefore, do not factor into the measured distance. \citet{Roth+2021} also noted that the brightest \pn in their sample is $\SI{0.11}{\mag}$ brighter than the second brightest \pn. While not really overluminous, they re-fit their data without this point and obtain a better fit with a larger distance modulus of $(m-M)=\SI[parse-numbers=false]{\uncertainty{29.87}{0.03}{0.05}}{\mag}$. 
Taking the catalogue from \citet{Roth+2021}, we run our fitting algorithm and measure a larger distance modulus of $(m-M)=\SI[parse-numbers=false]{\uncertainty{29.97}{0.02}{0.05}}{\mag}$ (also using the zero point of $M^*=-4.53$ that they adopted in their paper), indicating that the fitting algorithm and especially the treatment of the uncertainties plays an important role (as \citet{Roth+2021} convolved the luminosity function with the photometric error). This galaxy also has a number of \trgb distances \citep[e.g.][]{2009AJ....138..332J} and they are all in excellent agreement. \medskip

\textbf{\galaxyname{NGC}{1087}}: lies in the constellation Cetus. The $15$~\pn{}e that are brighter than our completeness limit yield a distance modulus of $(m-M)=\SI[parse-numbers=false]{\getvalue{mag/NGC1087}}{\mag}$ ($D=\SI[parse-numbers=false]{\getvalue{Mpc/NGC1087}}{\mega\parsec}$). This is in agreement with the distances derived from the numerical action method (NAM) or by group affiliation (see Figure~\ref{fig:NGC1087_distances}). \medskip

\textbf{\galaxyname{NGC}{1300}}: is located in the constellation Eridanus and part of the eponymous cluster. Our source detection picks up many extended sources that are associated with very significant $\HA$ emission. Even though this galaxy does not have an active nucleus, we mask the central region and spiral arms to get rid of those objects. We measure a distance modulus of $(m-M)=\SI[parse-numbers=false]{\getvalue{mag/NGC1300}}{\mag}$ ($D=\SI[parse-numbers=false]{\getvalue{Mpc/NGC1300}}{\mega\parsec}$). Our value is considerably larger than the existing literature values (see Figure~\ref{fig:NGC1300_distances}). However the small sample size of only 17 \pn{}e means that this value is more of an upper limit.
\medskip

\textbf{\galaxyname{NGC}{1365}}: is also known as the \emph{Great Barred Spiral Galaxy} and is part of the Fornax cluster. This galaxy has an active nucleus and many of the detected objects do not have a clean radial profile, but are instead dominated by noise. From $29$~\pn{}e we derive a distance modulus of $(m-M)=\SI[parse-numbers=false]{\getvalue{mag/NGC1365}}{\mag}$ ($D=\SI[parse-numbers=false]{\getvalue{Mpc/NGC1365}}{\mega\parsec}$). This galaxy also has a number of other reliable distance estimates (Figure~\ref{fig:NGC1365_distances}). Both the two most recent Cepheid distances and the \trgb distance agree well with our measured distance, while another \trgb and Cepheid distance indicate a larger distance. \medskip
    
\textbf{\galaxyname{NGC}{1385}}: we only detect $11$~\pn{}e that are brighter than our completeness limit. Most of them are not isolated and have other $\OIII$ emitting sources nearby, which makes the background subtraction more challenging. The derived distance modulus of $(m-M)=\SI[parse-numbers=false]{\getvalue{mag/NGC1385}}{\mag}$ ($D=\SI[parse-numbers=false]{\getvalue{Mpc/NGC1385}}{\mega\parsec}$) is therefore subject to a large uncertainty and should be considered an upper limit for the distance. The value is in good agreement with existing Tully--Fisher distances but shows a large discrepancy to the distance derived with the numerical action method (see Figure~\ref{fig:NGC1385_distances}). $9$~\snr{}s are potentially misclassified as \pn{}e. None of them fall at the bright end and their inclusion has the opposite effect of increasing the measured distance by $\SI{0.08}{\mag}$. \medskip
    
\textbf{\galaxyname{NGC}{1433}}: also known as \emph{Miltron's Galaxy}, is a barred spiral galaxy with an active nucleus. We detect $90$~\pn{}e that are all isolated with regular radial profile. This is a bit surprising, as the resulting distance modulus of $(m-M)=\SI[parse-numbers=false]{\getvalue{mag/NGC1433}}{\mag}$ ($D=\SI[parse-numbers=false]{\getvalue{Mpc/NGC1433}}{\mega\parsec}$) places it at the far end of our sample where we usually do not find a clean sample. Our measured distance is also significantly larger than all existing literature distances, including a \trgb distance that was measured from LEGUS--\textit{HST} data \citep{2018ApJS..235...23S}. We downloaded and reduced the same set of \textit{HST} data, using the technique outlined in \citet{Anand+2021}, however we are unable to derive a distance based on the color-magnitude diagram we obtain. The data are relatively shallow and a possible explanation for the discrepancy is that \citet{2018ApJS..235...23S} measured the tip of the AGB and not the tip of the red giant branch. \medskip

\textbf{\galaxyname{NGC}{1512}}: is a barred spiral galaxy that is interacting with the nearby galaxy \galaxyname{NGC}{1510}. We find $43$~\pn{}e, from which we measure $(m-M)=\SI[parse-numbers=false]{\getvalue{mag/NGC1512}}{\mag}$ ($D=\SI[parse-numbers=false]{\getvalue{Mpc/NGC1512}}{\mega\parsec}$). Similar to \galaxyname{NGC}{1433}, our derived distance is considerably larger than all existing literature distances (see Figure~\ref{fig:NGC1512_distances}), including again the \trgb distance from LEGUS--\textit{HST} \citep{2018ApJS..235...23S}. \medskip

\textbf{\galaxyname{NGC}{1566}}: is located in the constellation Dorado. We identify $28$~\pn{}e, but one of them is visibly separated from the rest of the luminosity function (by $\SI{1}{\mag}$) and including it results in a very poor fit. We therefore mark the brightest \pn as overluminous and redo the analysis and get a much better fit ($D_\mathrm{max}=0.135$, $p\textrm{-value} =0.66$) with a distance modulus of $(m-M)=\SI[parse-numbers=false]{\getvalue{mag/NGC1566}}{\mag}$ ($D=\SI[parse-numbers=false]{\getvalue{Mpc/NGC1566}}{\mega\parsec}$). This is considerably smaller than existing literature values based on the group affiliation \citep[][see Figure~\ref{fig:NGC1566_distances}]{2017ApJ...843...16K}. Five \snr{}s could be misclassifed as \pn{}e. The brightest among them is as bright as the overluminous \pn and is therefore also excluded. The four remaining \snr{}s do not alter the measured distance.\medskip

\textbf{\galaxyname{NGC}{1672}}: is a barred spiral galaxy with an active nucleus (Seyfert type~2). There are two objects that are significantly brighter than the rest (one \pn and one \snr). Even when we exclude them, the resulting fit is still rather poor with $D_\mathrm{max}=0.296$, $p\textrm{-value}=0.05$ and yields a distance modulus of $(m-M)=\SI[parse-numbers=false]{\getvalue{mag/NGC1672}}{\mag}$ ($D=\SI[parse-numbers=false]{\getvalue{Mpc/NGC1672}}{\mega\parsec}$). The measured value falls between the values derived from the NAM and those derived with Tully--Fisher measurements (see Figure~\ref{fig:NGC1672_distances}). \snr{}s do not impact the measured distance. \medskip

\textbf{\galaxyname{NGC}{2835}}: this is one of two galaxies where our completeness limit estimation yields a smaller value of $\SI{27}{\mag}$. However, we do not see a drop in the luminosity function until $\SI{27.5}{\mag}$ (see Figure~\ref{fig:all_galaxies_PNLF}) and, since the number of detected \pn{}e is already small, we decided to include objects up to $\SI{27.5}{\mag}$ in our sample. From the $27$~\pn{}e that match this criterion, we measure a distance modulus of $(m-M)=\SI[parse-numbers=false]{\getvalue{mag/NGC2835}}{\mag}$ ($D=\SI[parse-numbers=false]{\getvalue{Mpc/NGC2835}}{\mega\parsec}$) (we also measure $\SI{30.56}{\mag}$ from the sample with the smaller completeness limit). Existing Tully--Fisher distances arrive at a smaller distance, but a recent \trgb distance from \citet{Anand+2021} is in good agreement with our value (see Figure~\ref{fig:NGC2835_distances}).\medskip

\textbf{\galaxyname{NGC}{3351}}:  also known as \galaxyname{M}{95}, is a barred spiral galaxy located in the Leo constellation that has a characteristic ring. This is one of the few galaxies in our sample with an existing \pnlf distance. \citet{Ciardullo+2002} measured a distance modulus of $(m-M)=\uncertainty{30.05}{0.08}{0.16}$. We recover the six \pn{}e that fall in our field of view, however two of them are classified differently. We classify their 5\textsuperscript{th} brightest \pn as a \snr and their 6\textsuperscript{th} brightest object is classified as an \HII region. This galaxy has the second largest number of \pn detections in our sample and we find an excellent fit with a distance modulus of $(m-M)=\SI[parse-numbers=false]{\getvalue{mag/NGC3351}}{\mag}$ ($D=\SI[parse-numbers=false]{\getvalue{Mpc/NGC3351}}{\mega\parsec}$). This is significantly larger than the previous \pnlf distance and also larger than most of the existing \trgb  \citep[e.g.,][]{2007ApJ...661..815R,2018ApJS..235...23S} and Cepheid \citep[e.g.,][]{2006ApJS..165..108S,Freedman+2001} distances. There are however also a \trgb distance \citep{2004ApJ...608...42S} that is in good agreement and a Cepheid distance \citep{2006A&A...452..423P} that is even larger (see Figure~\ref{fig:NGC3351_distances}).\medskip

\textbf{\galaxyname{NGC}{3627}}:  also known as \galaxyname{M}{66}, can also be found in the constellation Leo and is part of the Leo Triplet. This is the second galaxy from \citet{Ciardullo+2002} that is also in our sample, but only one of the \pn{}e that was detected in that paper falls within our field of view. We recover this object and measure the same $m_{\OIII}$ magnitude but classify it as a \snr. This object is the 35\textsuperscript{th} brightest \pn in the catalogue of \citet{Ciardullo+2002}. Therefore, this misclassification will hardly impact the result. Their brighter \pn{}e all fall outside of our field of view and, hence, we can not make any statements about them. In our \pn sample, this object would constitute one of the brightest object. Including this object would have a minuscule effect on our result and decrease the measured distance modulus by $\SI{0.02}{\mag}$. The distance modulus that we measure from the clean \pn{}e sample is $(m-M)=\SI[parse-numbers=false]{\getvalue{mag/NGC3627}}{\mag}$ ($D=\SI[parse-numbers=false]{\getvalue{Mpc/NGC3627}}{\mega\parsec}$). This is in good agreement with the existing \pnlf distance of $(m-M)=\uncertainty{29.99}{0.07}{0.08}$ or other recent \trgb distances \citep[e.g.][see Figure~\ref{fig:NGC3627_distances}]{Anand+2021,2009AJ....138..332J}. There are $8$ potential \snr contaminants, one of which is overluminous. If we exclude this object, the measured distance is not impacted by the other $7$ \snr{}s.\medskip

\textbf{\galaxyname{NGC}{4254}}:  also known as \galaxyname{M}{99}, is one of four galaxies in our sample that are part of the Virgo cluster. $45$~objects constitute our initial \pn{}e catalogue. However, a large number of objects are surrounded by strong $\HA$ emission and three objects are rejected because of their irregular radial profile. Despite this, the remaining $42$ \pn{}e provided an excellent fit with $D_\mathrm{max}=0.071$, $p\textrm{-value}=0.982$ and yield a distance modulus of $(m-M)=\SI[parse-numbers=false]{\getvalue{mag/NGC4254}}{\mag}$ ($D=\SI[parse-numbers=false]{\getvalue{Mpc/NGC4254}}{\mega\parsec}$). This is smaller than all previously published distances. Among them are two distances based on the type~II supernova SN1986I \citep{2006ApJ...645..841N,2009ApJ...694.1067P} and a number of Tully--Fisher estimates. They all indicate a larger distance modulus around $(m-M)=\SI{30.75}{\mag}$ (see Figure~\ref{fig:NGC4254_distances}). The Hubble velocity of this galaxy is $\SI{2400}{\km\per\second}$ \citep{Springob+2005} and also implies a significant larger distance. The inclusion of $22$~misclassified \snr{}s does not affect the measured distance (one \snr is removed because it is clearly overluminous).
\medskip

\textbf{\galaxyname{NGC}{4303}}: also known as \galaxyname{M}{61}, is also part of the Virgo Cluster and an active (Seyfert type~2) barred spiral galaxy. We detect $19$ \pn{}e from which we measure $(m-M)=\SI[parse-numbers=false]{\getvalue{mag/NGC4303}}{\mag}$ ($D=\SI[parse-numbers=false]{\getvalue{NGC4303}}{\mega\parsec}$), but the fit is not good with $D_\mathrm{max}=0.326$, $p\textrm{-value}=0.03$. This galaxy did not have a good distance estimate, only a number of type II supernovae that do not agree on one distance as well as some uncertain Tully--Fisher estimates. Eight \snr could be misclassified as \pn{}e. One of them is clearly overluminous ($\SI{1.2}{\mag}$ below the next three objects and $\SI{1.8}{\mag}$ below the rest). The next three brightest objects are also \snr{}s and they are $\SI{0.6}{\mag}$ brighter than the bright end of the \pnlf. While there is a gap in the luminosity function, their inclusion improves the fit ($D_\mathrm{max}=0.243$, $p\textrm{-value}=0.227$) and they are therefore retained in the sample. With them, the measured distance is decreased by $\SI{0.48}{\mag}$.\medskip

\textbf{\galaxyname{NGC}{4321}}: also known as \galaxyname{M}{100} is located in the Virgo cluster. It is a low ionisation nuclear emission region (LINER) galaxy and we mask out the centre and the spiral arms. We get a decent fit ($D_\mathrm{max}=0.103$, $p\textrm{-value}=0.50$) and measure a distance modulus of $(m-M)=\SI[parse-numbers=false]{\getvalue{mag/NGC4321}}{\mag}$ ($D=\SI[parse-numbers=false]{\getvalue{Mpc/NGC4321}}{\mega\parsec}$) from $62$~\pn{}e. \galaxyname{NGC}{4321} has existing literature distances from Cepheids \citep{2008A&ARv..15..289T,Freedman+2001} and a recent \trgb distance from \citet{Anand+2021}. They all agree with our reported value. Out of $11$ \snr{}s, one falls $\SI{0.6}{\mag}$ below the bright end cutoff. With this object, the measured distance modulus decreases to $\SI[parse-numbers=false]{\uncertainty{30.78}{0.08}{0.11}}{\mag}$. If we only include the other $10$ \snr{}s, the distance modulus is very similar to the one derived from the clean \pn sample ($\SI[parse-numbers=false]{\uncertainty{31.053}{0.06}{0.09}}{\mag}$). However the fit is not bad enough ($D_\mathrm{max}=0.108$, $p\textrm{-value}=0.336$), to exclude this one object as overluminous.\medskip
                      
\textbf{\galaxyname{NGC}{4535}}: is the fourth galaxy in our sample that is part of the Virgo Cluster. From $53$~\pn{} we measure $(m-M)=\SI[parse-numbers=false]{\getvalue{mag/NGC4535}}{\mag}$ ($\SI[parse-numbers=false]{\getvalue{Mpc/NGC4535}}{\mega\parsec}$). This is in good agreement to the Cepheid distance from \citet{2006A&A...452..423P} who reported a value of $\SI{31.6}{\mag}$, but significantly bigger than other Cepheid distances \citep[e.g.~][$\SI{30.85+-0.05}{\mag}$]{Freedman+2001}. This is the only galaxy for which we do not find any potentially misclassified \snr{}s. \\
Noteworthy is the large range of distances that we obtained for the four Virgo galaxies in our sample (ranging from $\SI{9.86}{\mega\parsec}$ for \galaxyname{NGC}{4254} to $\SI{19.29}{\mega\parsec}$ for \galaxyname{NGC}{4535}). Among all galaxies in the cluster, there are a few published distances that are smaller than our value for \galaxyname{NGC}{4254}, but neither of them is coming from \trgb, Cepheid or \pnlf measurements.\medskip

\textbf{\galaxyname{NGC}{5068}}: is located in the Virgo constellation. This is the nearest and also the least massive galaxy in our sample. We find $58$~\pn{}e and our fit yields a distance modulus of $(m-M)=\SI[parse-numbers=false]{\getvalue{mag/NGC5068}}{\mag}$ ($\SI[parse-numbers=false]{\getvalue{Mpc/NGC5068}}{\mega\parsec}$). This is another galaxies with an existing \pnlf distance from \citet{Herrmann+2008}. Unfortunately, only two of their \pn are in the central region that is covered by our observations. We only recover one of them and the $\OIII$ line map does not show any emission at the location of the second one. The data from \citet{Herrmann+2008} indicate a larger distance of $\SI[parse-numbers=false]{\uncertainty{28.85}{0.09}{0.16}}{\mag}$. Because of the lower metallicity of this galaxy, they decide to use a fainter $M^*$ and obtain the reported distance of $(m-M)=\uncertainty{28.68}{0.08}{0.18}$. We do not apply such a correction here, as this would even further decrease our measured distance, which is already smaller than the existing \pnlf and \trgb distances. Within the uncertainties, our distance is in good agreement to the \trgb distances from \citet{Anand+2021} and \citet{2017MNRAS.469L.113K}. While there is a large number of \snr{}s~(22), their inclusion does not alter the measured distance.\medskip

\textbf{\galaxyname{NGC}{7496}}: is a barred spiral galaxy with Seyfert type~2 activity, located in the constellation Grus. PHANGS--MUSE covered this galaxies with only $3$~pointings. Due to the active core we mask the central part and arms, leading to the detection of only $14$~\pn{}e and $2$~\snr{}s that could be misclassified. One \pn{}e was excluded as an overluminous object. Previous distance estimates were all based on the Tully--Fisher method, with a large scatter between the different studies. Our distance modulus of $(m-M)=\SI[parse-numbers=false]{\getvalue{mag/NGC7496}}{\mag}$ ($\SI[parse-numbers=false]{\getvalue{Mpc/NGC7496}}{\mega\parsec}$) is in good agreement to NAM distance modulus of $(m-M)=\SI{31.36\pm0.33}{\mag}$. Both of the \snr{}s fall slightly below the bright and decreases the measured distance by $\SI{0.60}{\mag}$. The few data points that we have for this galaxy makes it difficult to judge whether these two points belong to the sample or not.\medskip

\section{Discussion}
\label{sec:discussion}

\subsection{Comparison with literature distances}

\begin{figure*}
\centering
\includegraphics[width=\textwidth]{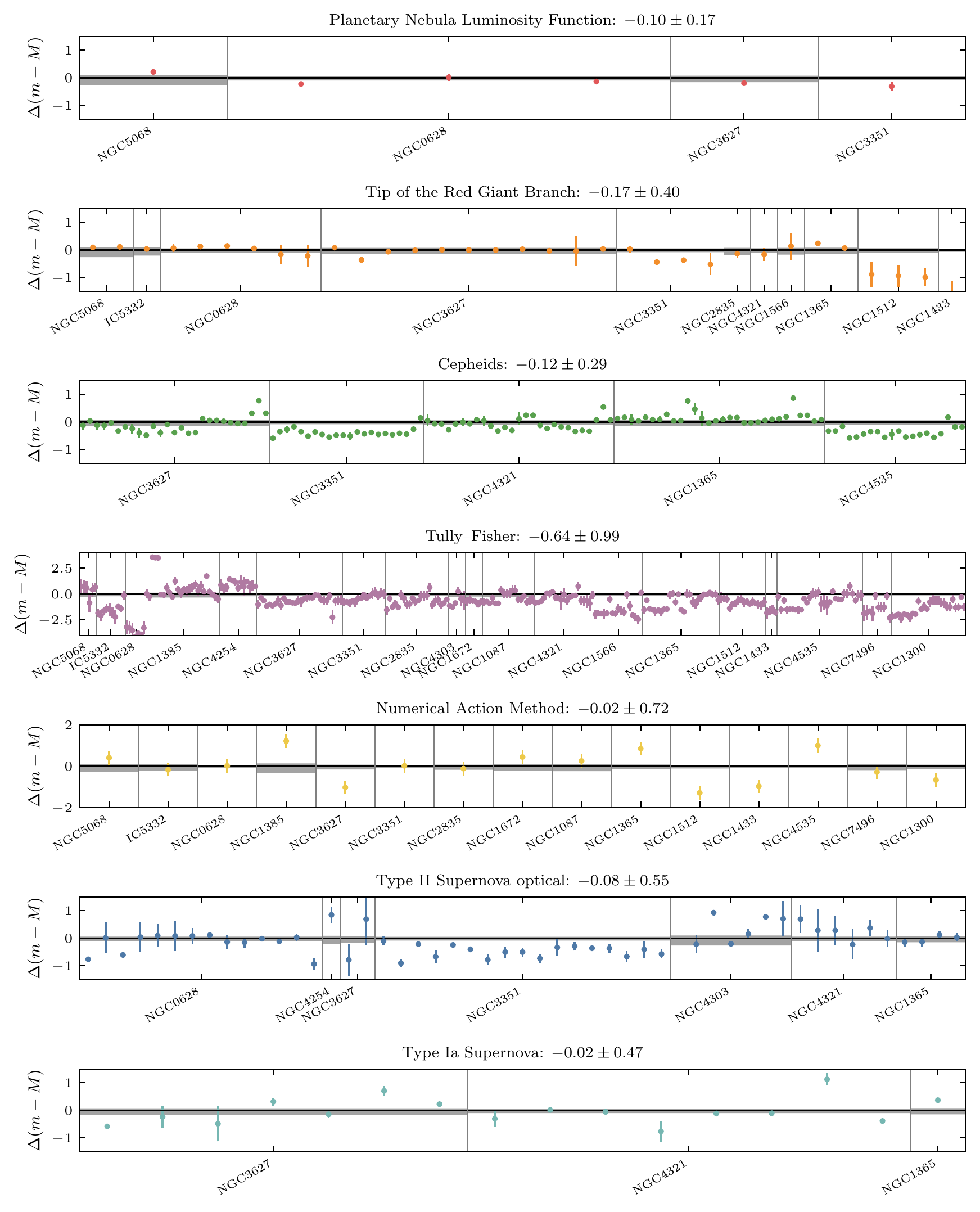}
\caption{Comparison with literature distances, taken from NED. Shown is the difference between our value and the literature $\Delta(m-M)=(m-M)_\mathrm{ref}-(m-M)_\mathrm{PHANGS}$. The galaxies are sorted by distance and the points for each galaxy are sorted by year of publication. If more than one value was published for a source, each value is shown separately. The uncertainties of our measurement is indicated by the grey shaded area.}
\label{fig:literature_distances}
\end{figure*}

To benchmark our results, we compare them to published distances. We have literature distances for all galaxies, for some galaxies from a myriad of different methods (see Appendix~\ref{sec:literature_distances}). For the four galaxies with existing \pnlf distances, we find excellent agreement with less than $1\sigma$ difference for \galaxyname{NGC}{0628} (derived from the same data) and \galaxyname{NGC}{3627}, while the previous \pnlf distance to \galaxyname{NGC}{5068} is slightly larger ($2\sigma$) and the previous \pnlf distance to \galaxyname{NGC}{3351} is significantly smaller ($3.8\sigma$).

Ten galaxies  also have \trgb distances and with the exception of \galaxyname{NGC}{1433} and \galaxyname{NGC}{1512}, they are in good agreement with our values. Furthermore, we have five galaxies with Cepheid distance. The distances derived with this method often vary between studies and our results always fall within the range of published distances. \pnlf distances have historically shown a systematic offset to distances measured from surface brightness fluctuations (\sbf) \citep{Ciardullo+2002}. However, as there are no elliptical galaxies in the PHANGS sample, we have no \sbf distances to compare with. To check for systematic offsets between our \pnlf distances and other methods, we plot the difference $\Delta (m-M) = (m-M)_\mathrm{\mathrm{ref}}-(m-M)_\mathrm{\mathrm{PHANGS}}$ between a number of commonly used distance indicators and our derived distance modulus in Figure~\ref{fig:literature_distances}. The plot shows generally good agreement between the different methods, but there is a slight systematic offset as our \pnlf distances are on average larger than the literature distances across all methods. However the offsets are generally small and always smaller than the scatter. The rather large scatter compared to the literature \trgb distances mostly stems from the discrepancies with the distances to \galaxyname{NGC}{1433} and \galaxyname{NGC}{1512}. In Section~\ref{sec:results} we provided a possible explanation for this and if we exclude them from the comparison, the \trgb method shows an excellent agreement with $\Delta (m-M)=\SI{-0.04\pm0.18}{\mag}$.

The galaxies that previously lacked a good distance estimate (i.e.~from the \pnlf, \trgb or Cepheids) are \galaxyname{NGC}{1087}, \galaxyname{NGC}{1300}, \galaxyname{NGC}{1385}, \galaxyname{NGC}{1672}, \galaxyname{NGC}{4254}, \galaxyname{NGC}{4303} and \galaxyname{NGC}{7496}. Unfortunately, these are also the galaxies with the fewest \pn{} detections, which in general leads to larger errors on the measured distances. However they are usually still smaller than what is available in the literature. 

\subsection{Zero point of the \pnlf}
\label{sec:zeropoint}

The single most important input to the \pnlf is the zero point. For our analysis, we assume a constant value across all galaxies, but theoretical models \citep{Dopita+1992} and observations \citep{Ciardullo+2002} indicate an increase of $M^*$ at low metallicities. The change can be described by a quadratic formula,\footnote{Adopted from \citet{Ciardullo+2002} but with $12+\log (\mathrm{O}/\mathrm{H})_\odot$ from \citet{Asplund+2009} instead of \citet{Grevesse+1996}.}
\begin{equation}\label{eq:zeropoint}
\Delta M^* = 0.928 \, [\mathrm{O}/\mathrm{H}]^2 - 0.109 \, [\mathrm{O}/\mathrm{H}] + 0.004~,
\end{equation}
with the solar oxygen abundance $12+\log (\mathrm{O}/\mathrm{H})_\odot=8.69$ from \citet{Asplund+2009}. We can compare our measured distances to literature distances to re-examine this relation. For this, we use the distances in Table~\ref{tbl:sample} that were curated by \citet{Anand+2021}. We only use the subsample that is based on the \trgb (8~galaxies). To first approximation the change in the zero point is given by $\Delta M^*\approx(m-M)_\mathrm{\pnlf} - (m-M)_\mathrm{ref}$. We use this as a starting point in our \pnlf fit and vary $M^*$ until we reproduce the literature distance. \\
One distinct advantage of \ifu observations is that one can derive the \pnlf distance and the metallicity from the same data. For our data, the gas-phase oxygen abundance is measured from \HII regions (based on the S\nobreakdash-\hspace{0pt}calibration from \citealt{Pilyugin+2016}) by \citet{Kreckel+2019} and \citet{Santoro+2021}. We use their linear abundance gradient fits to calculate the metallicity at the mean radius of the \pn{}e (see Table~\ref{tbl:sample}). In contrast, the metallicities that were used in \citet{Ciardullo+2002} were compiled by \citet{Ferrarese+2000b} from different literature sources \citep[most of them are based on the strong line method calibration by][]{Dopita+1986}. The galaxies that overlap with our sample (\galaxyname{NGC}{3351} and \galaxyname{NGC}{3627}) show a large discrepancy of ${\sim}\SI{0.7}{\dex}$ between the PHANGS and literature metallicity (considerably more than what could be explained by differences in the value for the solar oxygen abundance). More recent abundances from \citet{Pilyugin+2014} and \citet{ToribioSanCipriano+2017} (for the SMC and LMC) show much better agreement with our sample (see top panel of Figure~\ref{fig:zeropoint}). This result shows that such analysis is severely hampered by uncertainties in the metallicity prescriptions and greatly benefits from the homogeneous sample that \ifu observations can provide.

Our sample lacks low metallicity galaxies to constrain the decrease at low metallicities and \citet{Ciardullo+2012} found that for galaxies with metallicities above the LMC, the zero point is roughly constant. Even though we can not test the metallicity dependence, it is still worth to re-examine the zero point. Fitting a constant line to our own data (see lower panel of Figure~\ref{fig:zeropoint}) we find a zero point of $\uncertainty{-4.469}{0.049}{0.025}\, \si{\mag}$. The fit is heavily skewed by \galaxyname{NGC}{3351} where our distance is more than $3\sigma$ larger than the \trgb distance. We therefore use a mixture model for outlier pruning \citep{ForemanMackey+2014} with which  we measure a zero point of $\uncertainty{-4.542}{0.103}{0.059}\, \si{\mag}$. This value is in excellent agreement to the value that \citet{Ciardullo+2012} derived with the help of \trgb distances. Because our sample only consists of 8 points, one of which is rejected by the outlier pruning, we prefer to use the initial adopted value of $M^*=-4.47$ for the PNLF distances computed in this paper. Given the growing databases of galaxies with both \trgb distances and optical IFU data in the archives, this topic is ripe for future work to revisit this question using larger samples of galaxies.

\begin{figure}
    \centering
    \includegraphics{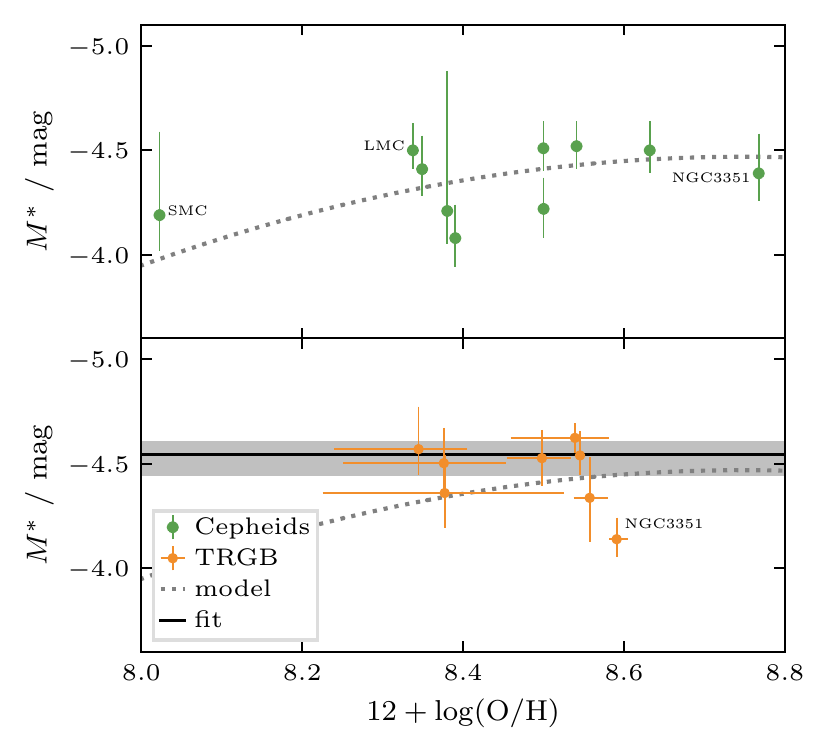}
    \caption{Determining the zero point of the \pnlf. The points in the top panel are based on Cepheid distances from \citet{Ciardullo+2002} and abundances from \citet{Pilyugin+2014}. The points in the lower panel are based on our measured distances in combination with existing \trgb distances from Table~\ref{tbl:sample}. The metallicity is determined by taking the value of the \HII region metallicity gradient \citep{Kreckel+2019,Santoro+2021} at the mean galactic radius of the \pn{}e. The error bars in $x$-direction represent the minimum and maximum abundance for the \pn{}e in our sample. The dotted curve shows the theoretical dependence from Equation~\ref{eq:zeropoint} and the solid line is a fit with a constant line (with the $1\sigma$ interval in grey). Overall our sample is consistent with a constant zero point.}
    \label{fig:zeropoint}
\end{figure}

\subsection{Contamination with supernova remnants}

Since \citet{Kreckel+2017} found that \snr{}s that are misclassified as \pn{}e can alter the distances measured from the \pnlf, the question stood whether narrowband observations are sufficient for \pn studies. In order for \snr{}s to bias the results, they must meet a couple of criteria. First of all, they must be classified as \pn{} by narrowband studies. As we see in Figure~\ref{fig:NGC628_emission_line_diagnostics}, most \snr{}s are in a regime where they would be classified as \HII regions without the additional diagnostics from the $\SII$ line. Second, the contamination must occur at the bright end of the luminosity function. The way the fitting of the \pnlf works, only the brightest objects have a noticeable contribution to the outcome. Third, the contaminant must not be too much brighter than the bright end cut-off. When a single object is much brighter than the main sample, it makes the fit worse and is therefore easily excluded as being overluminous. Hence, either a single \snr close to the bright end or a number of \snr{}s that are fainter than the bright end are needed to change the distance.

In almost all galaxies in our sample, we find at least one \snr that could be classified as a \pn without the $\SII$ criteria. However, only in the case of three galaxies do the misclassified objects have a significant impact the measured distance. This is the case for \galaxyname{NGC}{4303} ($\SI{-0.48}{\mag}$), \galaxyname{NGC}{4321} ($\SI{-0.33}{\mag}$) and \galaxyname{NGC}{7496} ($\SI{-0.60}{\mag}$). In the case of \galaxyname{NGC}{4303}, three \snr{}s fall $\SI{0.5}{\mag}$ below the cutoff, however they do not worsen the fit, making it difficult to justify their exclusion. \galaxyname{NGC}{4321} has one \snr that is clearly separated from the rest of the sample, but the goodness of the fit is not impacted. Finally for \galaxyname{NGC}{7496} the sample is too small to reliably exclude the two slightly overluminous \snr{}s.

Directly related to this is the challenge in accurately determining the flux emitted by an object. The largest uncertainty arises from background contamination. The $\OIII$ line fluxes are relatively robust because there is little to no background emission. However, this is not the case for $\HA$. The background can have a significant contribution and so correct background subtraction is crucial. As shown in Figure~\ref{fig:NGC0628_fluxes_comparison}, different approaches to the background subtraction can lead to vastly different results for the $\HA$ fluxes. If not done properly, wrong fluxes can alter the line ratios and hence the classification of objects.

\subsection{Limitations of IFU observations to the \pnlf}

The \pnlf is rarely applied to galaxies beyond $\SI{20}{\mega \parsec}$, since with narrowband imaging, the expected flux from a single \pn at larger distances is similar to the background. Recent attempts by \citet{Ventimiglia+2011}, \citet{Arnaboldi+2013}, or \citet{Roth+2021} have pushed the limits of what is possible by using multi-slit imaging spectroscopy or \ifu spectroscopy. Here, we make a rough estimate on the maximum distance that can be measured from MUSE with exposure times similar to our observations ($\SI{43}{\min}$). We require a certain number of \pn detections (from our experience ${\gtrsim}20$) in order to measure a reliable distance . However, it is difficult to predict the number of \pn{}e that one can expect to observe in a given galaxy, as it depends on a number of parameters. It relies foremost on the underlying stellar population. More stars means more \pn{}e. To compare the number of detected \pn{}e between galaxies, it is useful to define the \emph{luminosity-specific planetary nebula number}\footnote{this definition is different from the $\alpha_{2.5}$ that is common in the literature \citep[e.g.,][]{Hartke+2017} in that we only use the number of \pn{}e that are brighter than our completeness limit and not within a fixed range above the bright end cut off.},
\begin{equation}
    \alpha = \log_{10} \frac{N_\pn}{L_\mathrm{bol}}~,
\end{equation}
where $N_\pn$ is the number of detected \pn{}e above our completeness limit and $L_\mathrm{bol}$ is the bolometric luminosity of the surveyed area. The bolometric luminosity of the survey area is estimated from a simulated $V$-band image via a bolometric correction. We use $f_\lambda = \SI{363.1e-11}{\erg \per \second \per \cm \squared \per \angstrom}$ \citep{Bessell+1998} as the zero point to convert fluxes into magnitudes and then calculate the bolometric luminosity as
\begin{equation}
    L_\mathrm{bol} = 10^{-0.4(M_V-4.79)} \, 10^{-0.4(\mathrm{BC}_V+0.07)} \, \si{\Lsun}~,
\end{equation}
with $\mathrm{BC}_V=\SI{-0.85}{\mag}$ \citep[from][]{Buzzoni+2006}. Since we also have stellar mass maps, we could also utilise those instead, but due to its prevalence in the literature, we opt to use $L_\mathrm{bol}$. Both yield similar results. 

Because all galaxies have the same observed completeness limit (for \galaxyname{NGC}{2835} and \galaxyname{NGC}{3627} we extrapolate the number of \pn{}e to match the other galaxies), the part of the \pnlf that we observe decreases with distance. This reduces $N_\pn$ while $L_\mathrm{bol}$ remains unaffected and, hence, we expect a decrease in $\alpha$ as can be seen in Figure~\ref{fig:specificPNnumber}. As evident in Figure~\ref{fig:specificPNnumber}, compared to \galaxyname{NGC}{0628}, more than half of the galaxies in our sample have fewer \pn{}e than expected. We find up to a factor of~$7$ difference in the number of \pn{}e. A possible explanation is a decrease of $\alpha$ with metallicity. But our galaxies are all similar and only show small variations in metallicity ($\SI{0.24}{\dex}$). The observed change is too large to be explained by the models of \citet{Buzzoni+2006}. However, a recent study by \citet{GalandeAnta+2021} also finds comparable differences.

We can use this relation to calculate the farthest distance at which we are able to obtain a sufficient number of \pn{}e to measure a reliable distance. For each galaxy, we use the previously measured $L_\mathrm{bol}$ to calculate the $\alpha$ that corresponds to $N_\pn=20$. We can then compute the distance that corresponds to this value based on the sampling of the \pnlf. Based on this analysis, we find the largest possible distance that could be measured for a galaxy in our sample is ${\sim}\SI{26}{\mega\parsec}$.

To some degree, the number of detections should also decrease with resolution as fainter objects start to blend in with their background at coarser resolution. However, \citet{Kreckel+2017} showed that this only becomes an issue above $\SI{100}{\parsec}$ physical resolution. With the typical seeing that we achieve with MUSE ($\SI{0.60}{\arcsec}$ with AO and $\SI{0.76}{\arcsec}$ without), this corresponds to a distance of $\SI{34.38}{\mega\parsec}$ and $\SI{27.14}{\mega\parsec}$, respectively, both above what we could measure. This means that this issue should only concerns less powerful \ifu instruments with a coarser spatial resolution and indeed we do not see any obvious connection between the resolution and the number of detections in our data.

\begin{figure}
\centering
\includegraphics{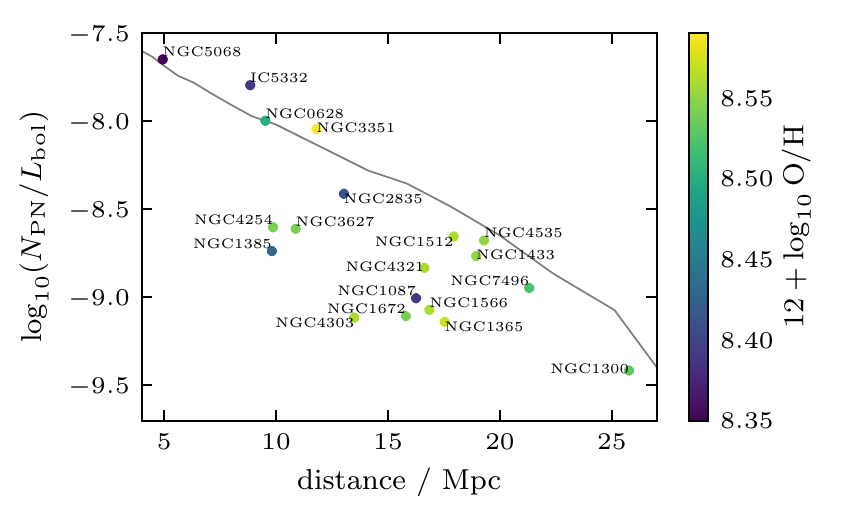}
\caption{Luminosity-specific planetary nebula number as a function of distance. The number of detections decreases with distance because we sample a smaller range of the \pnlf. The grey line uses \galaxyname{NGC}{0628} as a reference and shows the expected decrease of $\alpha$ based on the sampled part of the \pnlf.}
\label{fig:specificPNnumber}
\end{figure}

\section{Conclusions}
\label{sec:conclusion}

We use VLT/MUSE observations obtained by the PHANGS collaboration to identify 899 \pn{}e across 19 nearby galaxies. We then calculate new \pnlf distances for each galaxy (see Table~\ref{tbl:distances}), 15 of which did not have a \pnlf distance before. This significantly increases the number of galaxies with \pnlf distances, which previously stood at around~$70$. Fourteen galaxies achieve an uncertainty better than $\SI{10}{\percent}$ and six galaxies better than $\SI{5}{\percent}$. 

When comparing with other studies, we find good agreement of the $\OIII$ line fluxes but significant variations in the $\HA$ line fluxes. This can be traced back to different techniques for background subtraction (unlike $\OIII$, $\HA$ has a significant background contribution). This highlights the difficulty of the background subtraction which can have a major impact on the classification and, thereby, ultimately also on the measured distance.

The findings of \citet{Kreckel+2017} raised questions about the reliability of narrowband observations for \pn{}e studies, with the possibility of misclassified \snr{}s biasing the measured distance. While we can not reproduce this issue in the particular case of \galaxyname{NGC}{0628}, we find three other galaxies where \snr{}s can impact the measured distance. For the other galaxies, the changes are insignificant and smaller than the uncertainties. Most misclassified \snr{s} fall in the regime of \HII regions, and hence, do not contribute to the measured distance. The few \snr{}s that are classified as \pn{}e often are at the faint end of the luminosity function where their contribution is negligible.

With metallicities derived from the same MUSE data and \trgb distances from the literature, we revisit the calibration of the \pnlf zero point. Our sample is consistent with a constant zero point and we find a value of $\uncertainty{-4.542}{0.103}{0.059}\, \si{\mag}$.

Because they do not require dedicated observations, \ifu surveys can potentially multiply the number of galaxies with \pnlf distances in the near future. A growing number of spectral imaging surveys that utilise VLT/MUSE like MAD \citep{denBrok+2020} or TIMER \citep{Gadotti+2019}, or CFHT/SITELLE like SIGNALS \citep{RousseauNepton+2019} fill the archives with a plethora of suitable observations that could be harnessed to measure \pnlf distances.

\section*{Acknowledgements}

We thank the anonymous referee for the helpful comments that improved this work. This work was carried out as part of the PHANGS collaboration. Based on observations collected at the European Southern Observatory under ESO programmes 1100.B-0651, 095.C-0473, and 094.C-0623 (PHANGS--MUSE; PI Schinnerer), as well as 094.B-0321 (MAGNUM; PI Marconi), 099.B-0242, 0100.B-0116, 098.B-0551 (MAD; PI Carollo) and 097.B-0640 (TIMER; PI Gadotti). 
FS and KK gratefully acknowledges funding from the German Research Foundation (DFG) in the form of an Emmy Noether Research Group (grant number KR4598/2-1, PI Kreckel). 
GA acknowledges support from the IPAC Visiting Graduate Fellowship programme and from an award from the Space Telescope Science Institute in support of programme SNAP-15922. 
FS, ES and TGW acknowledge funding from the ERC under the European Union’s Horizon 2020 research and innovation programme (grant agreement no. 694343).
ATB and FB would like to acknowledge funding from the European Research Council (ERC) under the European Union’s Horizon 2020 research and innovation programme (grant agreement No.726384/Empire).
SCOG and RSK acknowledge support from the DFG via SFB 881 ‘The Milky Way System’ (project-ID 138713538; subprojects A1, B1, B2, and B8) and from the Heidelberg cluster of excellence EXC 2181-390900948 ‘STRUCTURES: A unifying approach to emergent phenomena in the physical world, mathematics, and complex data’, funded by the German Excellence Strategy. RSK also thanks for support from the European Research Council via the ERC Synergy Grant 'ECOGAL -- Understanding our Galactic ecosystem: From the disk of the Milky Way to the formation sites of stars and planets' (contract number 855130).
JMDK gratefully acknowledges funding from the Deutsche Forschungsgemeinschaft (DFG, German Research Foundation) through an Emmy Noether Research Group (grant number KR4801/1-1) and the DFG Sachbeihilfe (grant number KR4801/2-1), as well as from the European Research Council (ERC) under the European Union’s Horizon 2020 research and innovation programme via the ERC Starting Grant MUSTANG (grant agreement number 714907). 
ER acknowledges the support of the Natural Sciences and Engineering Research Council of Canada (NSERC), funding reference number RGPIN-2017-03987.
EW acknowledges support from the DFG via SFB 881 ‘The Milky Way System’ (project-ID 138713538; subproject P2). 

This research has made use of the NASA/IPAC Extragalactic Database (NED) which is operated by the Jet Propulsion Laboratory, California Institute of Technology, under contract with NASA. It also made use of a number of python packages, namely \textsc{photutils}, an Astropy package for detection and photometry of astronomical sources \citep{Photutils+2019} as well as the main \textsc{astropy} package \citep{Astropy+2013,Astropy+2018}, \textsc{numpy} \citep{Harris+2020} and \textsc{matplotlib} \citep{Hunter+2007}.


\section*{Data availability}
The MUSE data underlying this article are presented in \citet{Emsellem+2021} and are available at the ESO archive. The catalogue with the planetary nebulae identifications is available in the online supplementary material of the journal. The code for this project is available at \url{https://github.com/fschmnn/pnlf}.


\bibliographystyle{mnras}
\bibliography{paper,distances} 



\appendix
\section{Comparison with literature distances}\label{sec:literature_distances}

In Figures~\ref{fig:IC5332_distances} to \ref{fig:NGC7496_distances}, we compare our measured distances with literature values. The data for the plots were taken from the NASA/IPAC Extragalactic Database (NED).\footnote{\url{https://ned.ipac.caltech.edu/Library/Distances/}} If a source published more than one distance, the mean of the published distances is used with the uncertainties added in quadrature (those values are marked with a diamond). Only the five most recent publications are considered for each method. The Numerical action method (NAM) distances were obtained by \citet{Anand+2021} with the online tool\footnote{\url{http://edd.ifa.hawaii.edu/NAMcalculator/}} by \citet{2020AJ....159...67K} and \citet{2017ApJ...850..207S}.\\
We use the following abbreviations in the figures: Planetary Nebula Luminosity Function (\pnlf), Tip of the Red Giant Branch (\trgb), Type Ia Supernova (SNIa), SNII optical (SNII), Numerical action method (NAM), distance to galaxy group (Group), Tully--Fisher (TF), Tully estimate (TE), Brightest Stars (BS), Gravitational Stability Gaseous Disk (GSGD), Disk Stability (DS), Infra-Red Astronomical Satellite (IRAS), Ring Diameter (RD) and Statistical (Stat). 

\begin{figure}
\centering
\input{fig/IC5332_distances.pgf}
\caption[Caption for LOF (not used but otherwise I get an error with the footnote)]{Same as Figure~\ref{fig:NGC0628_distances}, but for \galaxyname{IC}{5332}. We measure a distance modulus of $(m-M)=\SI[parse-numbers=false]{\getvalue{mag/IC5332}}{\mag}$ ($D=\SI[parse-numbers=false]{\getvalue{Mpc/IC5332}}{\mega\parsec}$).}
\label{fig:IC5332_distances}
\end{figure}

\begin{figure}
\centering
\input{fig/NGC1087_distances.pgf}
\caption{Same as Figure~\ref{fig:NGC0628_distances}, but for \galaxyname{NGC}{1087}. We measure a distance modulus of $(m-M)=\SI[parse-numbers=false]{\getvalue{mag/NGC1087}}{\mag}$ ($D=\SI[parse-numbers=false]{\getvalue{Mpc/NGC1087}}{\mega\parsec}$).}
\label{fig:NGC1087_distances}
\end{figure}

\begin{figure}
\centering
\input{fig/NGC1300_distances.pgf}
\caption[Caption for LOF (not used but otherwise I get an error with the footnote)]{Same as Figure~\ref{fig:NGC0628_distances}, but for \galaxyname{NGC}{1300}. We measure a distance modulus of $(m-M)=\SI[parse-numbers=false]{\getvalue{mag/NGC1300}}{\mag}$ ($D=\SI[parse-numbers=false]{\getvalue{Mpc/NGC1300}}{\mega\parsec}$).}
\label{fig:NGC1300_distances}
\end{figure}

\begin{figure}
\centering
\input{fig/NGC1365_distances.pgf}
\caption[Caption for LOF (not used but otherwise I get an error with the footnote)]{Same as Figure~\ref{fig:NGC0628_distances}, but for \galaxyname{NGC}{1365}. We measure a distance modulus of $(m-M)=\SI[parse-numbers=false]{\getvalue{mag/NGC1365}}{\mag}$ ($D=\SI[parse-numbers=false]{\getvalue{Mpc/NGC1365}}{\mega\parsec}$).}
\label{fig:NGC1365_distances}
\end{figure}

\begin{figure}
\centering
\input{fig/NGC1385_distances.pgf}
\caption[Caption for LOF (not used but otherwise I get an error with the footnote)]{Same as Figure~\ref{fig:NGC0628_distances}, but for \galaxyname{NGC}{1385}. We measure a distance modulus of $(m-M)=\SI[parse-numbers=false]{\getvalue{mag/NGC1385}}{\mag}$ ($D=\SI[parse-numbers=false]{\getvalue{Mpc/NGC1385}}{\mega\parsec}$).}
\label{fig:NGC1385_distances}
\end{figure}

\begin{figure}
\centering
\input{fig/NGC1433_distances.pgf}
\caption[Caption for LOF (not used but otherwise I get an error with the footnote)]{Same as Figure~\ref{fig:NGC0628_distances}, but for \galaxyname{NGC}{1433}. We measure a distance modulus of $(m-M)=\SI[parse-numbers=false]{\getvalue{mag/NGC1433}}{\mag}$ ($D=\SI[parse-numbers=false]{\getvalue{Mpc/NGC1433}}{\mega\parsec}$).}
\label{fig:NGC1433_distances}
\end{figure}

\begin{figure}
\centering
\input{fig/NGC1512_distances.pgf}
\caption[Caption for LOF (not used but otherwise I get an error with the footnote)]{Same as Figure~\ref{fig:NGC0628_distances}, but for \galaxyname{NGC}{1512}. We measure a distance modulus of $(m-M)=\SI[parse-numbers=false]{\getvalue{mag/NGC1512}}{\mag}$ ($D=\SI[parse-numbers=false]{\getvalue{Mpc/NGC1512}}{\mega\parsec}$).}
\label{fig:NGC1512_distances}
\end{figure}

\begin{figure}
\centering
\input{fig/NGC1566_distances.pgf}
\caption[Caption for LOF (not used but otherwise I get an error with the footnote)]{Same as Figure~\ref{fig:NGC0628_distances}, but for \galaxyname{NGC}{1566}. We measure a distance modulus of $(m-M)=\SI[parse-numbers=false]{\getvalue{mag/NGC1566}}{\mag}$ ($D=\SI[parse-numbers=false]{\getvalue{Mpc/NGC1566}}{\mega\parsec}$).}
\label{fig:NGC1566_distances}
\end{figure}

\begin{figure}
\centering
\input{fig/NGC1672_distances.pgf}
\caption[Caption for LOF (not used but otherwise I get an error with the footnote)]{Same as Figure~\ref{fig:NGC0628_distances}, but for \galaxyname{NGC}{1672}. We measure a distance modulus of $(m-M)=\SI[parse-numbers=false]{\getvalue{mag/NGC1672}}{\mag}$ ($D=\SI[parse-numbers=false]{\getvalue{Mpc/NGC1672}}{\mega\parsec}$).}
\label{fig:NGC1672_distances}
\end{figure}

\begin{figure}
\centering
\input{fig/NGC2835_distances.pgf}
\caption[Caption for LOF (not used but otherwise I get an error with the footnote)]{Same as Figure~\ref{fig:NGC0628_distances}, but for \galaxyname{NGC}{2835}. We measure a distance modulus of $(m-M)=\SI[parse-numbers=false]{\getvalue{mag/NGC2835}}{\mag}$ ($D=\SI[parse-numbers=false]{\getvalue{Mpc/NGC2835}}{\mega\parsec}$).}
\label{fig:NGC2835_distances}
\end{figure}

\begin{figure}
\centering
\input{fig/NGC3351_distances.pgf}
\caption[Caption for LOF (not used but otherwise I get an error with the footnote)]{Same as Figure~\ref{fig:NGC0628_distances}, but for \galaxyname{NGC}{3351} (\galaxyname{M}{95}). We measure a distance modulus of $(m-M)=\SI[parse-numbers=false]{\getvalue{mag/NGC3351}}{\mag}$ ($D=\SI[parse-numbers=false]{\getvalue{Mpc/NGC3351}}{\mega\parsec}$).}
\label{fig:NGC3351_distances}
\end{figure}

\begin{figure}
\centering
\input{fig/NGC3627_distances.pgf}
\caption[Caption for LOF (not used but otherwise I get an error with the footnote)]{Same as Figure~\ref{fig:NGC0628_distances}, but for \galaxyname{NGC}{3627} (\galaxyname{M}{66}). We measure a distance modulus of $(m-M)=\SI[parse-numbers=false]{\getvalue{mag/NGC3627}}{\mag}$ ($D=\SI[parse-numbers=false]{\getvalue{Mpc/NGC3627}}{\mega\parsec}$).}
\label{fig:NGC3627_distances}
\end{figure}

\begin{figure}
\centering
\input{fig/NGC4254_distances.pgf}
\caption[Caption for LOF (not used but otherwise I get an error with the footnote)]{Same as Figure~\ref{fig:NGC0628_distances}, but for \galaxyname{NGC}{4254} (\galaxyname{M}{99}). We measure a distance modulus of $(m-M)=\SI[parse-numbers=false]{\getvalue{mag/NGC4254}}{\mag}$ ($D=\SI[parse-numbers=false]{\getvalue{Mpc/NGC4254}}{\mega\parsec}$).}
\label{fig:NGC4254_distances}
\end{figure}

\begin{figure}
\centering
\input{fig/NGC4303_distances.pgf}
\caption[Caption for LOF (not used but otherwise I get an error with the footnote)]{Same as Figure~\ref{fig:NGC0628_distances}, but for \galaxyname{NGC}{4303} (\galaxyname{M}{61}). We measure a distance modulus of $(m-M)=\SI[parse-numbers=false]{\getvalue{mag/NGC4303}}{\mag}$ ($D=\SI[parse-numbers=false]{\getvalue{Mpc/NGC4303}}{\mega\parsec}$).}
\label{fig:NGC4303_distances}
\end{figure}

\begin{figure}
\centering
\input{fig/NGC4321_distances.pgf}
\caption[Caption for LOF (not used but otherwise I get an error with the footnote)]{Same as Figure~\ref{fig:NGC0628_distances}, but for \galaxyname{NGC}{4321} (\galaxyname{M}{100}). We measure a distance modulus of $(m-M)=\SI[parse-numbers=false]{\getvalue{mag/NGC4321}}{\mag}$ ($D=\SI[parse-numbers=false]{\getvalue{Mpc/NGC4321}}{\mega\parsec}$).}
\label{fig:NGC4321_distances}
\end{figure}

\begin{figure}
\centering
\input{fig/NGC4535_distances.pgf}
\caption[Caption for LOF (not used but otherwise I get an error with the footnote)]{Same as Figure~\ref{fig:NGC0628_distances}, but for \galaxyname{NGC}{4535}. We measure a distance modulus of $(m-M)=\SI[parse-numbers=false]{\getvalue{mag/NGC4535}}{\mag}$ ($D=\SI[parse-numbers=false]{\getvalue{Mpc/NGC4535}}{\mega\parsec}$).}
\label{fig:NGC4535_distances}
\end{figure}

\begin{figure}
\centering
\input{fig/NGC5068_distances.pgf}
\caption[Caption for LOF (not used but otherwise I get an error with the footnote)]{Same as Figure~\ref{fig:NGC0628_distances}, but for \galaxyname{NGC}{5068}. We measure a distance modulus of $(m-M)=\SI[parse-numbers=false]{\getvalue{mag/NGC5068}}{\mag}$ ($D=\SI[parse-numbers=false]{\getvalue{Mpc/NGC5068}}{\mega\parsec}$).}
\label{fig:NGC5068_distances}
\end{figure}

\begin{figure}
\centering
\input{fig/NGC7496_distances.pgf}
\caption[Caption for LOF (not used but otherwise I get an error with the footnote)]{Same as Figure~\ref{fig:NGC0628_distances}, but for \galaxyname{NGC}{7496}. We measure a distance modulus of $(m-M)=\SI[parse-numbers=false]{\getvalue{mag/NGC7496}}{\mag}$ ($D=\SI[parse-numbers=false]{\getvalue{Mpc/NGC7496}}{\mega\parsec}$).}
\label{fig:NGC7496_distances}
\end{figure}

\section{Extinction correction}\label{sec:extinction}

To quantify if we can ignore the effects of internal extinction (dust within the target galaxy, not the circumstellar extinction of the \pn itself), we create a synthetic luminosity function by sampling from Equation~\ref{eq:pnlf}. To half of the sample we randomly add extinction. Then we use our fitting algorithm to measure the distance to the sampled data. Figure~\ref{fig:extinction_pnlf} shows the \pnlf without extinction and with extinction. As shown in Appendix~\ref{sec:pnlf_stat}, the ability to derive a reliable distance depends on the sample size and the range of the \pnlf that we sample. For a sample size of 20 \pn{}e, the measured distance is unaffected for $A_{5007}=\SI{0.1}{\mag}$, but can be significantly overestimated for $A_{5007}=\SI{1}{\mag}$. Once the sample gets larger, the impact decreases and for a sample size of 100 \pn{}e, our algorithm is able to derive the correct distances from the compound sample with any $A_{5007}$. 

As Figure~\ref{fig:extinction_pnlf} shows, the composition of a luminosity function should bend the observed luminosity function below the fitted function. Since we do not observe this behaviour in our observed sample, this further indicates that our sample is not significantly reddened. 

\begin{figure}
\centering
\includegraphics[width=\columnwidth]{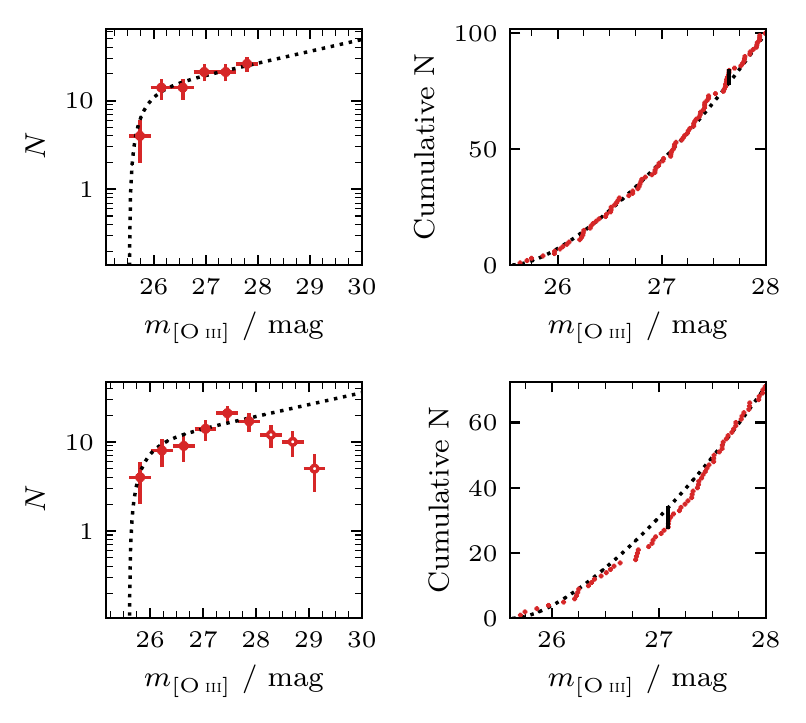}
\caption{We synthesis a \pnlf with 100 points. Half of the points are randomly chosen and an extinction of $A_{5007} = \SI{1}{\mag}$ is applied. The top panels show the full sample without extinction and the bottom panels show the compound sample with the extinction applied. The left panels show the binned \pnlf and the right panels show the cumulative \pnlf.} 
\label{fig:extinction_pnlf}
\end{figure}

\section{Precision of the \pnlf}\label{sec:pnlf_stat}

The fit of the \pnlf is dominated by the brightest \pn{}e. Once we observe a \pn{}e with apparent magnitude $m_{\OIII}$, the galaxy can not be further away than $(m-M)_\text{max}=m_{\OIII}-M^*$, where $M^*$ is the zero point of the \pnlf (it can be slightly further away due to the way we include the uncertainties with Equation~\ref{eq:pnlf_with_error}). Due to the exponential tail of the \pnlf, the fainter \pn{}e push the function to the right (e.g.~towards larger distances). Hence if we only observe the bright end, we usually underestimate the distance.

\citet{Jacoby1997} investigated the effects of sample size by measuring the \pnlf from different sized subsets from the \pn{}e detected in \galaxyname{M}{87}. To test the reliability of the \pnlf, we do something similar with synthetic data. We vary the sample size from 20 to 150 objects and the completeness limit from \SIrange{26.5}{28}{\mag}. For each step we sample 1000 luminosity functions with a fixed distance modulus of $(m-M)=\SI{30}{\mag}$. The later means that the part of the \pnlf that we sample varies from \SIrange{1.0}{2.5}{\mag}. The result can be seen in Figure~\ref{fig:sample_PNLF}. The precision increases with sample size, showing that even from a small sample of 20 \pn{}e, one can achieve a precision better than $\sim\SI{5}{\percent}$. For a brighter completeness limit, the sample is concentrated at the steep cut off, and hence the distance is very well constrained. For fainter completeness limits, this part is sampled more sparsely which leads to a lower precision of the measured distance.

\begin{figure*}
\centering
\includegraphics[width=\textwidth]{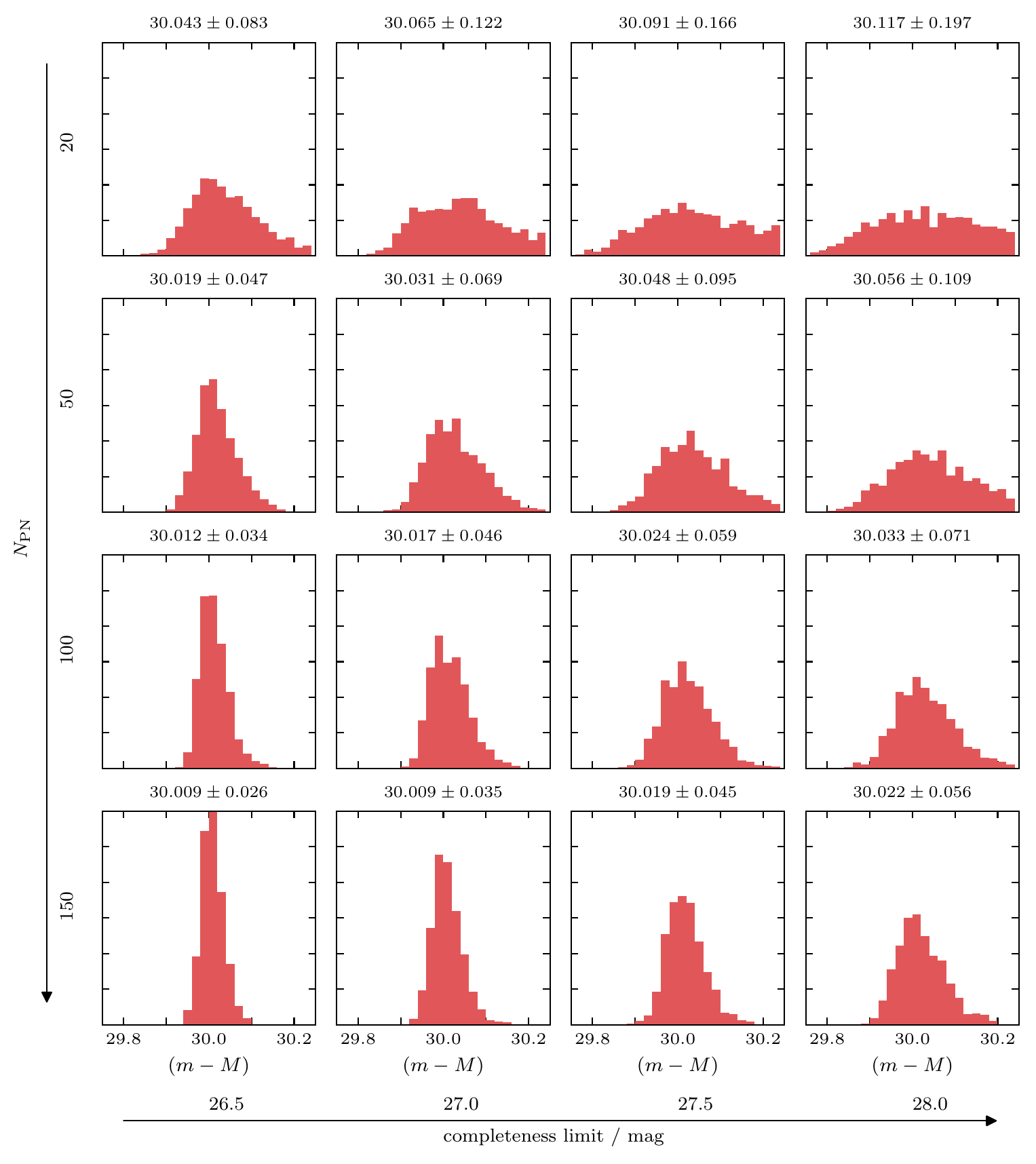}
\caption{A demonstration of the precision that can be achieved with the \pnlf, depending on the sample size and part of the \pnlf that is sampled. We sample the \pnlf for a distance modulus of $(m-M)=\SI{30}{\mag}$. The sample size $N_\mathrm{PN}$ increases from 20, 50, 100 to 150 and the completeness limit increases from \SIrange{26.5}{28}{\mag} in $\SI{0.5}{\mag}$ increments (meaning that we sample \SIrange{1.0}{2.5}{\mag} of the \pnlf). The precision increases with sample size and even a small sample of 20 \pn{}e is enough to achieve a precision around $\SI{5}{\percent}$. When we only sample a small part of the luminosity function, the sample is concentrated at the bright cut-off and the distance is well constrained. With a fainter completeness limit, the sample is distributed more and the more sparsely sampled cut-off yields a less precise distance.} 
\label{fig:sample_PNLF}
\end{figure*}


\vspace{4mm}

\noindent {\it
$^{1}$Astronomisches Rechen-Institut, Zentrum f\"{u}r Astronomie der Universit\"{a}t Heidelberg, M\"{o}nchhofstra\ss e 12-14, 69120 Heidelberg, Germany\\
$^{2}$Space Telescope Science Institute, 3700 San Martin Drive, Baltimore, MD 21218, USA\\
$^{3}$Institute for Astronomy, University of Hawaii, 2680 Woodlawn Drive, Honolulu, HI 96822, USA\\
$^{4}$Observatories of the Carnegie Institution for Science, Pasadena, CA 91101, USA \\
$^{5}$Departamento de Astronom\'{i}a, Universidad de Chile, Casilla 36-D, Santiago, Chile\\
$^{6}$Max-Planck-Institut f\"{u}r Astronomie, K\"{o}nigstuhl 17, D-69117 Heidelberg, Germany\\
$^{7}$IPAC, California Institute of Technology, Pasadena, CA 91125, USA\\
$^{8}$Argelander-Institut f\"{u}r Astronomie, Universit\"{a}t Bonn, Auf dem H\"{u}gel 71, 53121, Bonn, Germany
$^{9}$Universit\"{a}t Heidelberg, Zentrum f\"{u}r Astronomie, Institut f\"{u}r theoretische Astrophysik, Albert-Ueberle-Str. 2, 69120 Heidelberg, Germany \\
$^{10}$Research School of Astronomy and Astrophysics, Australian National University, Canberra, ACT 2611, Australia\\
$^{11}$Universit\"{a}t Heidelberg, Interdisziplin\"{a}res Zentrum f\"{u}r Wissenschaftliches Rechnen, INF 205, 69120 Heidelberg, Germany \\
$^{12}4-183$ CCIS, University of Alberta, Edmonton, Alberta, Canada\\
$^{13}$Max-Planck-Institute for extraterrestrial Physics, Giessenbachstraße
1, D-85748 Garching, Germany\\
}

\bsp	
\label{lastpage}
\end{document}
